\documentclass[12pt]{article}
\usepackage{amsmath}
\usepackage{amssymb}
\numberwithin{equation}{section}

\title{Rational solutions to the Pfaff lattice
  \\ and Jack polynomials}

\author{
M. Adler\thanks{ Department of Mathematics, Brandeis University,
Waltham, Mass 02454, USA. E-mail: adler@brandeis.edu.  The support
of a National Science Foundation grant \# DMS-01-00782 is
gratefully acknowledged.}~~~~~~ V. B. Kuznetsov\thanks{Department
of Applied Mathematics, University of Leeds, Leeds LS2 9JT.
E-mail: V.B.Kuznetsov@leeds.ac.uk. The support of a FSR grant,
Universite de Louvain, Belgium, and from the EPSRC, UK, is
gratefully acknowledged. This work was done, while visiting the
University of Louvain and Brandeis University. }~~~~~~ P. van
Moerbeke\thanks{ Department of Mathematics, Universit\'e de
Louvain, 1348 Louvain-la-Neuve, Belgium and Brandeis University,
Waltham, Mass 02454, USA. E-mail: vanmoerbeke@geom.ucl.ac.be and
@brandeis.edu. The support of a National Science Foundation grant
\# DMS-01-00782, a Nato, a FNRS and a Francqui Foundation grant is
gratefully acknowledged.}}

\date{February 15, 2002}

\catcode`\@=11
\let\c@equation=\relax
\newcounter{equation}[subsection]

\catcode`\@=12

\newcommand{\MAT}[1]{\left(\begin{array}{*#1c}}
\newcommand{\mat}{\end{array}\right)}
\newcommand{\qed}{\leavevmode\unskip\nobreak\penalty200\hskip2pt\null
\nobreak\hfill\rule{1.1ex}{1.1ex}
\medbreak
}

\newcommand{\rg}{\rightarrow}

\newcommand{\DF}{\Longleftrightarrow}
\newcommand{\df}{\longleftrightarrow}
\newcommand{\pp}{\ldots}

\newcommand{\BC}{{\mathbb C}}

\newcommand{\BL}{{\mathbb L}}
\newcommand{\BS}{{\mathbb S}}
\newcommand{\BX}{{\mathbb X}}
\newcommand{\BY}{{\mathbb Y}}
\newcommand{\BZ}{{\mathbb Z}}

\newcommand{\iy}{\infty}
\newcommand{\pl}{\partial}
\newcommand{\al}{\alpha}
\newcommand{\tp}{\tilde\partial}
\newcommand{\gs}{{\bf s}}
\newcommand{\no}{\nonumber}
\newcommand{\oiint}{\frac{1}{(2\pi
i)^2}\oint\limits_{\iy}\!\!\oint\limits_{\iy}}

\newenvironment{proof}{\medskip\noindent{\it Proof:\/} }{\qed}
\newenvironment
        {remark}{\medskip\noindent\underline{\it Remark:\/} }{\medbreak}
\newenvironment
        {example}{\medskip\noindent\underline{\it Example:\/} }{\medbreak}

\newcommand{\la}{\langle}
\newcommand{\ra}{\rangle}

 \newcommand{\vr}{\varepsilon}
\newcommand{\sg}{\sigma}
\newcommand{\BR}{{\mathbb R}}
\newcommand{\lb}{\lambda}
\newcommand{\Lb}{\Lambda}

\newcommand{\diag}{\operatorname{diag}}

\def\be#1\ee{\begin{equation}#1\end{equation}}
\def\bea#1\eea{\begin{eqnarray}#1\end{eqnarray}}
\def\bean#1\eean{\begin{eqnarray*}#1\end{eqnarray*}}

\newtheorem{definition}{Definition}[section]
\newtheorem{theorem}[definition]{Theorem}
\newtheorem{lemma}[definition]{Lemma}
\newtheorem{corollary}[definition]{Corollary}
\newtheorem{proposition}[definition]{Proposition}

\catcode `!=11

\newdimen\squaresize
\newdimen\thickness
\newdimen\Thickness
\newdimen\ll! \newdimen \uu! \newdimen\dd! \newdimen \rr! \newdimen
\temp!

\def\sq!#1#2#3#4#5{%
\ll!=#1 \uu!=#2 \dd!=#3 \rr!=#4
\setbox0=\hbox{%
 \temp!=\squaresize\advance\temp! by .5\uu!
 \rlap{\kern -.5\ll!
 \vbox{\hrule height \temp! width#1 depth .5\dd!}}%
 \temp!=\squaresize\advance\temp! by -.5\uu!
 \rlap{\raise\temp!
 \vbox{\hrule height #2 width \squaresize}}%
 \rlap{\raise -.5\dd!
 \vbox{\hrule height #3 width \squaresize}}%
 \temp!=\squaresize\advance\temp! by .5\uu!
 \rlap{\kern \squaresize \kern-.5\rr!
 \vbox{\hrule height \temp! width#4 depth .5\dd!}}%
 \rlap{\kern .5\squaresize\raise .5\squaresize
 \vbox to 0pt{\vss\hbox to 0pt{\hss $#5$\hss}\vss}}%
}
 \ht0=0pt \dp0=0pt \box0
}

\def\vsq!#1#2#3#4#5\endvsq!{\vbox to \squaresize{\hrule
width\squaresize height 0pt%
\vss\sq!{#1}{#2}{#3}{#4}{#5}}}

\newdimen \LL! \newdimen \UU! \newdimen \DD! \newdimen \RR!

\def\vvsq!{\futurelet\next\vvvsq!}
\def\vvvsq!{\relax
  \ifx     \next l\LL!=\Thickness \let\continue=\skipnexttoken!
  \else\ifx\next u\UU!=\Thickness \let\continue=\skipnexttoken!
  \else\ifx\next d\DD!=\Thickness \let\continue=\skipnexttoken!
  \else\ifx\next r\RR!=\Thickness \let\continue=\skipnexttoken!
  \else\def\continue{\vsq!\LL!\UU!\DD!\RR!}%
  \fi\fi\fi\fi
  \continue}

\def\skipnexttoken!#1{\vvsq!}

\def\place#1#2#3{\vbox to 0pt{\vss
\rlap{\kern#1\squaresize
  \raise#2\squaresize\hbox{$#3$}}
\vss}}

\def\Young#1{\LL!=\thickness \UU!=\thickness \DD! = \thickness \RR! =
\thickness \vbox{\smallskip\offinterlineskip \halign{&\vvsq! ##
\endvsq!\cr #1}}}

\catcode `!=12

\begin{document}
\maketitle

$$
\mbox{\bf To the memory of J\"urgen Moser}
$$

\vspace{.2cm}

\begin{abstract}

The finite Pfaff lattice is given by commuting Lax pairs involving
a finite matrix $L$ (zero above the first subdiagonal) and a
projection onto $Sp(N)$. The lattice admits solutions such that
the entries of the matrix $L$ are rational in the time parameters
$t_1,t_2,\ldots$, after conjugation by a diagonal matrix. The
sequence of polynomial $\tau$-functions, solving the problem,
belongs to an intriguing chain of subspaces of Schur polynomials,
associated to Young diagrams, dual with respect to a finite chain
of rectangles. Also, this sequence of $\tau$-functions is given
inductively by the action of a fixed vertex operator.

As examples, one such sequence is given by Jack polynomials for
rectangular Young diagrams, while another chain starts with any
two-column Jack polynomial.

\end{abstract}

\newpage

\tableofcontents

\section{Introduction}

 \noindent{\bf Self-dual partitions}:
  For positive integer $n$ and $n|k$, define the
following sets of partitions,
  \bean
  \BY&=& \{ \lb=(\lb_1,\lb_2,...),~~\lb_1\geq\lb_2\geq\dots\geq 0 \}\\
   \BY_k&=& \{ \lb \in \BY,~~ |\lb|=\sum \lb_i= k\}\\
\BY^{(n)}_k&=&\left\{\begin{array}{cr}
\lb=(\lb_1,\lb_2,...,\lb_n)\in \BY_k , & ~\hat\lb_1\leq n,
\\ &  \\
  \lb_i+\lb_{n+1-i}=\frac{2k}{n},&
 1\leq i\leq \left[ \frac{n+1}{2}  \right]
\end{array}
\right\}, \eean
with
$$
 \# \BY^{(n)}_k=\left(   {{\left[\frac{n}{2}+\frac{k}{n}\right] }
 \atop {\left[  \frac{n+1}{2}  \right] }} \right).
 $$
These are a few examples:

$$
\BY^{(4)}_8=\left\{\begin{array}{cccccc}
 \squaresize .4cm
\thickness .01cm \Thickness .07cm \Young{
 &&&\cr
 &\cr
 &\cr}

 ,&

 \squaresize .4cm \thickness .01cm \Thickness .07cm \Young{
 &&\cr
 &\cr
 &\cr
 \cr}

 ,&

 \squaresize .4cm \thickness .01cm \Thickness .07cm \Young{
 &&&\cr
 &&&\cr}

 ~,&

 \squaresize .4cm \thickness .01cm \Thickness .07cm \Young{
 &\cr
 &\cr
 &\cr
 &\cr}

 ~,&

 \squaresize .4cm
\thickness .01cm \Thickness .07cm \Young{
 &&&\cr
 &&\cr
 \cr}

 ~,&
  \squaresize .4cm
\thickness .01cm \Thickness .07cm \Young{
 &&\cr
 &&\cr
 \cr
 \cr}

 \end{array}
  \right\}
 $$

$$
\BY^{(3)}_6=\left\{\begin{array}{ccccc}

\squaresize .4cm \thickness .01cm \Thickness .07cm \Young{
 &&\cr
 &\cr
 \cr}

 ,&

\squaresize .4cm \thickness .01cm \Thickness .07cm \Young{
 &&&\cr
 &\cr}

 ,&

 \squaresize .4cm \thickness .01cm \Thickness .07cm \Young{
 &\cr
 &\cr
 &\cr}

 \end{array}  \right\}
. $$

\bigbreak

\noindent Let ${\bf s}_{\lb}(t):=\det (s_{\lb_{i}-i+j}(t))_{1\leq
i,j}$ be the Schur polynomials corresponding to $\lb$, with
 ${\bf s}_i(t)$ being the elementary Schur polynomials, defined by
$$\mbox{$\displaystyle{e^{\sum_1^{\iy}t_iz^i}=\sum_{i\geq 0}}{\bf
s}_i(t)z^i$ ~~with ~~${\bf s}_i(t)=0$ for $i<0$}.$$
The linear space
 $$
\BL_k^{(n)} :=\left\{ \sum_{\lb \in \BY_k^{(n)}} a_{\lb}{\bf
s}_{\lb}
  ~\Bigl|~ a_{\lb}\in \BC\right\}
  $$
will play an ubiquitous role in this work.

\bigbreak

 \noindent{\bf The finite Pfaff Lattice}: The $N\times N$
skew-symmetric matrices, \bea
   J&=&\left(
\begin{array}{cc@{}c@{}cc}
 &\boxed{\begin{array}{cc} 0 & 1 \\ -1 & 0 \end{array}} &&0& \\
 && \ddots
 &&\\
 &0&& \boxed{\begin{array}{cc} 0 & 1 \\ -1 & 0 \end{array}} & \\
 &&&& 
 \end{array}
 \right),~~~\mbox{for
\underline{$N$ even}}
 \nonumber\\&&\nonumber\\
  &=&\left(
\begin{array}{cc@{}c@{}cc}
 &\boxed{\begin{array}{cc} 0 & 1 \\ -1 & 0 \end{array}} &&0& \\
 && \ddots
  &&\\
 &0&& \boxed{\begin{array}{cc} 0 & 1 \\ -1 & 0 \end{array}} & \\
 &&&& 0
 \end{array}
 \right),~~~\mbox{for
\underline{$N$ odd}},
 \eea
satisfy
 \be J^2=
 \left\{\begin{array}{l}
 -I_N, \mbox{~for $N$ even}\\
  \\
  \left(
  \begin{array}{cc}
 -I_{N-1}&O\\
 O&0
  \end{array}
  \right), \mbox{~for $N$ odd.}
 \end{array}
  \right.
 \ee
Also consider the Lie algebra $\frak k$ of lower-triangular
matrices of the form
\be
{\frak k }=\left\{
\begin{array}{c}
 \left(
\begin{array}{c@{}c@{}ccc}
 & \boxed{\begin{array}{cc}
 a_1 & 0 \\ 0 & a_1 \end{array}} &&&O\\
 &&& \ddots   \\
  &*&& &\boxed{\begin{array}{cc}
  a_{[N/2]} & 0 \\ 0 & a_{[N/2]} \end{array}}
 \end{array}
 \right), \mbox{~for $N$ even}\\
  \\
\left(
\begin{array}{c@{}c@{}cccc}
  &&& \\
 & \boxed{\begin{array}{cc}
 a_1 & 0 \\ 0 & a_1 \end{array}} &&& O & \\
 &&& \ddots   \\
  &&& &\boxed{\begin{array}{cc}
  a_{[N/2]} & 0 \\ 0 & a_{[N/2]} \end{array}} &  \\
  &*&&& &\boxed{\begin{array}{c}
  a_{(N+1)/2}  \end{array}}
 \end{array}
 \right), \mbox{~for $N$ odd.}
 \end{array}
 \right.
 \ee
%
%
%
 For each $a\in gl(N)$, consider the decomposition\footnote{$a_{\pm}$ refers to
projection onto strictly upper (strictly lower)
triangular matrices, with all $2\times 2$ diagonal
blocks equal zero. $a_{0}$ refers to projection onto
the ``diagonal", consisting of $2\times 2$ blocks. }
 \bea
 a&=&(a)_{{\frak k}}
+(a)_{{\frak n}} \nonumber\\ &=&\pi_{{\frak k}}
a+\pi_{{\frak n}} a\nonumber\\
&=&\left((a_--J(a_+)^{\top}J)+\frac{1}{2}
(a_0-J(a_0)^{\top}J)\right)  \nonumber\\ &&~~~~~~~~~+
\left((a_++J(a_+)^{\top}J)+\frac{1}{2}(a_0+J(a_0)^{\top}
J)\right). \eea
For \underline{ $N$ even}, this corresponds to a Lie
algebra splitting, given by
 $$ gl(N)={\frak
k}+{\frak n}\left\{
\begin{array}{l}
{\frak k}=\{\mbox{lower-triangular matrices of the
form (1.0.3)}\}\\ {\frak n}=sp(N)=\{\mbox{$a$ such that
$Ja^{\top}J=a$}\},
\end{array}
\right. $$ \vspace{-1.3cm}\be \ee

For \underline{ $N$ odd}, this is merely a vector
space splitting
 $$ gl(N)={\frak
k}+{\frak n}\left\{
\begin{array}{l}
{\frak k}=\{\mbox{lower-triangular matrices of the form
(1.0.3)}\}\\ {\frak n}=\mbox{span} \{\pi_{\frak
n}(a)~~\mbox{with}~a \in gl(N)\}.
\end{array}
\right. $$ \vspace{-1.3cm}\be \ee

\vspace{1cm}

 The Pfaff lattice is defined on $N\times N$ matrices
$L$ of the form
 \bean
L&=&\left(\begin{array}{cccccccc} 0&1& & &&&& \\
 &-d_1 &a_1& & & & &O
\\
 & &d_1 &1  & & & &\\
 & &    &-d_2   &a_2 & & &\\
 & &    &       & d_2 &  & & \\
 & \ast &  &  &  &\ddots &a_{\frac{N-2}{2}} & \\ 
 &  &  &  &  & & -d_{\frac{N-2}{2}}& 1  \\
 &  &  &  &  & &  & 0
  \end{array}
\right) ,~~\mbox{for $N$
 even}
 \\
%
%
&=&\left(\begin{array}{cccccccc} 0&1& & &&&& \\
 &-d_1 &a_1& & & & &O
\\
 & &d_1 &1  & & & &\\
 & &    &-d_2   &a_2 & & &\\
 & &    &       & d_2 &  & & \\
 & \ast &  &  &  &\ddots &1 & \\ 
 &  &  &  &  & & -d_{\frac{N-1}{2}}& a_{\frac{N-1}{2}}  \\
 &  &  &  &  & &  & d_{\frac{N-1}{2}}
  \end{array}
\right)  ,~~\mbox{for $N$
 odd}.
 \eean  \vspace{-1.4cm}\be  \ee
 namely,
  \be
   \frac{\pl L}{\pl
t_{i}}=[-(L^i)_{\frak k},L].~~~~~~~\mbox{\bf (Pfaff
lattice) }
 \label{Pfaff Lattice}\ee

\bigbreak


%
\noindent Given arbitrary, but fixed parameters
 \be
b_0,...,b_{[\frac{N-2}{2}]} \in \BC, \label{b}
 \ee
  consider the skew-symmetric antidiagonal initial
condition,
  \bean m_N(0)&=&\left(
\begin{array}{cccccc}
 &O& & & &b_{\frac{N-2}{2}}\\
 & & & &\diagup& \\
 & & & b_0& & \\
& &-b_0& & & \\
 &\diagup& & & & \\
-b_{\frac{N-2}{2}}& & & &O&
\end{array}
\right)   ~~\mbox{,  for $N$ even,} \nonumber \\ &&
 \nonumber\\
   &=&\left(
\begin{array}{ccccccc}
 &&O& & & &b_{\frac{N-3}{2}}\\
 && & & &\diagup& \\
 && & & b_0& & \\
 && &0& & & \\
&& -b_0&& & & \\
 &\diagup&& & & & \\
-b_{\frac{N-3}{2}}&& & & &O&
\end{array}
\right)
  ~~\mbox{,  for $N$ odd}
   \eean   \vspace{-1.4cm} \be \ee
and its time evolution (respecting the skew-symmetry),
   \be
   m_{\ell}(t)=E_{\ell,N}(t) m_N(0)
    E^{\top}_{\ell,N}(t),
    \ee
where\footnote{$\Lb$ is the finite shift matrix
 $\Lb:= (\delta_{i,j-1})_{1 \leq i,j \leq N}$ and $(A)_{{1,\ldots , \ell }
 \atop {1,\ldots,
   N}}$ denotes the
 matrix formed by the first $\ell$ rows and first $N$ columns of $A$.}
 \be
  E_{\ell,N}(t):= \left( e^{\sum_1^{\iy} t_i
  \Lb^i}\right)_{{1,\ldots , \ell }\atop {1,\ldots,
   N}}.
   \ee
    The Pfaffian $pf ~ m_{\ell}(t)$ of the
    skew-symmetric matrix $m_{\ell}(t)$
     will play an
    important role in this paper.

\bigbreak

 \noindent {\bf Rational solutions to the Pfaff Lattice}:
\begin{theorem}
 Modulo conjugation by a $N\times N$ diagonal
 matrix $D(t)$ (see remark below),
  the finite Pfaff lattice
  $$
   \frac{\pl L}{\pl
t_{i}}=[-(L^i)_{\frak k},L].~~~~~~~\mbox{\bf (Pfaff
lattice) }
 \label{Pfaff Lattice}$$
  has rational solutions
   in $t_1,t_2,...$; i.e., the matrix
 \be
D^{-1}(t)L(t)D(t)=\tilde Q(t)\Lambda \tilde Q(t)^{-1}
\ee
  is rational in $t_1,t_2,...$, with
  $\tilde Q(t)$ a lower-triangular $N\times N$
 matrix with rational entries,
 obtained by Taylor
 expanding $\tau_{2n}(t-[z^{-1}])$ in $z^{-1}$,
  with $\tau_0=1$,
 \bea \tilde q_{2n}(t;z)&:=&\sum_{j=0}^{2n}
   \tilde Q_{2n+1,j+1}(t) z^j = z^{2n}
 \tau_{2n}(t-[z^{-1}])~~~\mbox{with}~
0\leq n\leq \left[\frac{N-1}{2}\right] \nonumber\\
 \tilde q_{2n+1}(t;z)&:=&\sum_{j=0}^{2n+1}
   \tilde Q_{2n+2,j+1}(t) z^j= z^{2n}
(z+\frac{\pl}{ \pl t_1})\tau_{2n}(t-[z^{-1}]) ,
\nonumber\\
 \eea
 with (see the definition of the $\BL$-space in the beginning of
 this section)
    \bea \tau_{\ell}
{(t)}&=&
 {pf}~\left( E_{\ell,N}(t) m_N(0)
    E^{\top}_{\ell,N}(t)\right)\nonumber\\
  &=&\sum_{\lb\in\BY^{(\ell)}_{\frac{\ell
(N-\ell)}{2}}}\left(\prod_1^{[\ell/2]}b_{\lb_
i-i+\ell-\left[\frac{N+1}{2}\right]} \right){\bf
s}_{\lb}(t),\qquad \mbox{for}~\left\{
\begin{array}{l} 0\leq \ell\leq N-1\\ \ell~\mbox{~even}
\end{array}\right.  \nonumber
\\
&\in & \BL^{(\ell)}_{\frac{\ell (N-\ell)}{2}}~.
\label{tau-functions}\eea
The polynomials $q_k=D_k\tilde q_k$ (in $z$) of degree $0\leq k
\leq N-1$ are ``skew-orthonormal" with respect to
 the skew inner-product $\la z^i,z^j\ra= m_{ij}(t)$,
i.e.,
\be \la q_i,q_j \ra= J_{ij},\ee
 and the $N$-vector $(q_0, \ldots, q_{N-1})^{\top}$ is an
 eigenvector for the matrix $L$, with
 modified boundary conditions. The fact that $Q_{2n,2n-1}=0$
defines the skew-orthogonal polynomials in a unique way, up to
$\pm1$.
\end{theorem}

 \example For $\ell=2$, we have
 \bea
\tau_2(t)&=&\sum_{i=0}^{\frac{N-2}{2}} b_i
 {\bf s}_{\frac{N-2}{2}+i,\frac{N-2}{2}-i}(t)
 ,~~\mbox{ for $N$ even,}\nonumber\\
 &=&
  \sum_{i=0}^{\frac{N-3}{2}} b_i
 {\bf s}_{\frac{N-1}{2}+i,\frac{N-3}{2}-i}(t)
  ,~~\mbox{ for $N$ odd.}\eea

  \remark
 \bean
\lefteqn{D(t)}
\\
 &=&\diag\left(\frac{1}{\sqrt{\tau_0\tau_2}},\frac{1}{\sqrt{\tau_0\tau_2}},
\frac{1}{\sqrt{\tau_2\tau_4}},\frac{1}{\sqrt{\tau_2\tau_4}},...,
 \frac{1}{\sqrt{\tau_{N-2}\tau_N}},\frac{1}
  {\sqrt{\tau_{N-2}
 \tau_N}}\right)~~~\mbox{for N even}\\
&=&\diag\left(\frac{1}{\sqrt{\tau_0\tau_2}},\frac{1}{\sqrt{\tau_0\tau_2}},
  ,...,
  \frac{1}{\sqrt{\tau_{N-3}\tau_{N-1}}},
 \frac{1}{\sqrt{\tau_{N-3}\tau_{N-1}}},\frac{1}
  {\sqrt{\tau_{N-1} }}\right) ~~~\mbox{for N odd}.
\eean

\bigbreak

\bigbreak

 \noindent{\bf Duality:} For the case of odd $N$, we can even define
$\tau_{\ell}(t)$ for odd $\ell$, by slightly deforming the initial
moment matrix $m_N(0)$. In section 6, we prove a duality between
these $\tau_{k}$'s for $k$ even and odd, as follows
 $$
  \tilde \tau_{\ell}(t)=
(-1)^{\ell(N-\ell)/2} ~ \left(\prod_0^{\frac{N-3}{2}}b_i\right)
\left( \left.\tau_{N-\ell}(-t)\right|_{b_i\rightarrow
 b_i^{-1}}\right),~~~\mbox{for $\ell$ odd}.
 $$

\vspace{1cm}


\noindent{\bf Fay identities}:

\begin{theorem}The sequence of functions
 \be \tau_{\ell}
{(t)}=\sum_{\lb\in\BY^{(\ell)}_{\frac{\ell
(N-\ell)}{2}}}\left(\prod_1^{[\ell/2]}b_{\lb_
i-i+\ell-\left[\frac{N+1}{2}\right]} \right){\bf
s}_{\lb}(t),\qquad
\begin{array}{l} 0\leq \ell\leq N-1\\ \ell~\mbox{~even}
\end{array},
\ee together with the ``boundary condition"
 \be
\tau_0=1~~\mbox{ and}~~ \left\{\begin{array}{ll}
\tau_N=\displaystyle{\prod_0^{\frac{N-2}{2}} b_i},&\mbox{for even $N$}\\
\tau_{N+1}=0
 ,&\mbox{for odd $N$,}
\end{array} \right.
 \ee
 satisfies the
 the ``differential Fay identity\footnote{Define the Wronskian
$\{f,g\}=\frac{\pl f}{\pl t_1}g- \frac{\pl g}{\pl t_1}f.$}'':
\begin{multline}
\{\tau_{2n}(t-[u]),\tau_{2n}(t-[v])\}
\\
{}+(u^{-1}-v^{-1})(\tau_{2n}(t-[u])\tau_{2n}(t-[v])
-\tau_{2n}(t)\tau_{2n}(t-[u]-[v]))
\\
=uv(u-v)\tau_{2n-2}(t-[u]-[v])\tau_{2n+2}(t).
\end{multline}

\end{theorem}

\bigbreak

\noindent{\bf Vertex operator constructions of the
rational solutions}:
%
%
Consider the vertex operator acting on functions
$f(t)$ of $t=(t_1,t_2,\dots)\in\BC^{\iy}$, namely
\be
X^{}(t;z)=e^{\sum_{1}^\iy t_iz^i} e^{-
\sum_{1}^\iy\frac{z^{-i}}{i}\frac{\pl}{\pl t_i}};
\label{vertex} \ee and the vector vertex operator
%
%
\be
\BX(t;z)=\Lambda^{\top}e^{\sum^{\iy}_1t_iz^i}e^{-\sum^{\iy}_1\frac{z^{-i}}{i}
\frac{\pl}{\pl t_i}}\chi(z),\label{vectorvertex} \ee
acting on vectors of functions
$F=(f_0(t),f_1(t),\dots)$, with
$\chi(z):=(z^i)_{i\geq0}$. Then the composition
$\BX(t;\lb)\BX(t;\mu)$ is a vertex operator for the
Pfaff lattice, i.e., for any $\tau$-vector
$=(\tau_0,\tau_2,\tau_4,\dots)$ of the Pfaff lattice,
$$ \tau(t)+a\BX(t;\mu)\BX(t;\lb)\tau(t),\quad a\in\BC
$$ is again a $\tau$-vector of the Pfaff lattice, or
coordinatewise $$
\tau_{2n}+a\left(1-\frac{\lb}{\mu}\right)\mu^{2n-1}\lb^{2n-2}e^{\sum
t_i(\lb^i+\mu^i)}\tau_{2n-2}(t-[\lb^{-1}]-[\mu^{-1}])
$$ provides a new sequence of Pfaff $\tau$-functions.

\bigbreak

In terms of the distributional weight, with the $b_i$ as in
(1.0.9),

$$
\rho_b(x):=\left\{
\begin{array}{l}
\rho_b^{(e)}(x)=\displaystyle{\sum_{i\geq 0}}b_i(x^{-i-1}-x^i),\mbox{~for $N$
even,}\\
\\
\rho_b^{(0)}(x)=x^{-1/2}\displaystyle{\sum_{i\geq
0}}b_i(x^{-i-1}-x^{i+1}),\mbox{~for $N$
 odd.}\\
\end{array}
\right. $$
 and
  \be \beta
:=\frac{N}{2}-\ell +1,\label{beta}
 \ee
  we define
the {\em integrated vertex operator}, in terms of the
vertex operator (\ref{vertex}),
%
\bean  Y_{\beta}(t)&:= & 
 \oiint X(t;y)X(t;z)\frac{\rho_b(y/
z)dy~dz}{z^2(yz)^{\beta}}
\eean
and the {\em integrated vector vertex operator}, in
terms of the (\ref{vectorvertex}),
\be
\BY_N(t)  = 
 \oiint
\BX(t;y)\BX(t;z)
 \frac{\rho_b(y/
z)dy~dz}{2(yz)^{N/2}z}. \ee In both cases, the double integral
around two contours about $\iy$ amounts to computing the
coefficient of $1/yz$.



%


\begin{theorem} For a given set of $b_i$, the sequence of
$\tau$-functions $\tau_0,\tau_2,\tau_4,\dots$, defined
in (\ref{tau-functions}), is generated by the vertex
operators $Y_p$; to be precise, inductively $$
Y_{\frac{N}{2}-\ell +1}\tau_{\ell -2}=\ell\tau_{\ell}.
$$
\end{theorem}

\begin{corollary} The vector of $\tau$-functions
$$ I=(I_0,I_2,I_4,\dots),\mbox{~with
$I_{\ell}=\left(\frac{\ell}{2}\right)!\tau_{\ell}$} $$
is a fixed point for the vertex operator $\BY_N$,
namely $$ (\BY_N I)_{\ell}=I_{\ell},\mbox{~for even
$\ell$.} $$
\end{corollary}

The rational solutions to the Pfaff lattice can be $q$-deformed;
this will be reported on at a later stage.

\bigbreak


\noindent{\bf Example 1: Rectangular Jack polynomials}

Jack polynomials are symmetric polynomials in the variables $x_i$,
which are orthogonal with respect to the inner-product
 $$ \la p_{\lb},p_{\mu}\ra
=\delta_{\lb\mu}(1^{m_1}2^{m_2}\ldots)m_1!m_2!\ldots
 \al^{ \lb_1^{\top}},
$$
where $m_i=m_i(\lb)$ is the number of times that $i$ appears in
the partition $\lb$ and where
 $$
  p_{\lb}(x_1,x_2,\ldots ):=
p_{\lb_1}p_{\lb_2}\cdots
 =\sum_i x_i^{\lb_1}
  \sum_i x_i^{\lb_2} \cdots
  .$$
 Precise
definitions and properties of Jack polynomials can be found in
\cite{MacDonald,Stanley,awata,KS1,KS2}.

\begin{proposition} When

\bean
 b_i&=&2i+1~~\mbox{for $N$ even}\\
    &=& 2i+2 ~~ \mbox{for $N$ odd},
    \eean
 then the $\tau_{2n}(t)$'s are Jack polynomials for
rectangular partitions \bean \tau_{2n}(t)&=&
 \sum_{\lb\in\BY^{(2n)}_{{n(N-2n)}}}\prod_1^{n}
  (k_i-k_{2n+1-i}
){\bf s}_{\lb}(t), \mbox{~where~} \left\{
\begin{array}{l}
k_i=\lb_i-i+2n\\ 0\leq 2n\leq N,
\end{array}
\right.\\ & & \\ &=&pf~m_{2n}(t)\\
&=&\frac{1}{n!}\int_{\BR^n}\Delta(z)^4\prod_{k=1}^ne^{2\displaystyle{\sum^{\iy}_
1}t_i z^i_k}\delta^{(N-2)}_{(z_k)}dz_k\\ & & \\
&=&J^{(1/2)}_{\lb}(x)\Bigg|_{t_i=\frac{1}{i}
\displaystyle{\sum_k}x^i_k}\quad\mbox{for~}\lb
=\underbrace{(N-2n,...,N-2n)}_n \eean
  where the $m_{2n}(t) $'s are the $2n\times 2n$
  upper-left hand corners of
  \be
{m_N(t)} = \left((j-i)\tilde{\bf s}_{N-i-j-1}
 \right)_{0\leq i,j\leq N-1}.
 \ee
 upon setting $\tilde{\bf s}_n(t):={\bf s}_n(2t)$.
\end{proposition}

\noindent{\bf Example 2: Two-row Jack polynomials}

\begin{proposition} For even $N$,
choosing\footnote{$(a)_k=\displaystyle{\frac{\Gamma(a+k)}{\Gamma(a)}}=a(a+1)...(a+k-
1)$}
\be
 \left\{\begin{array}{ll}
b_0=...=b_{\frac{p}{2}-1}=0\\
\\
b_{\frac{p}{2}+k}=\frac{(1-\al)_k(p+1)_k}{k!(\al+p+1)_k},\mbox{~~for~~}
k=0,...,\frac{N-2-p}{2},
  \end{array}\right.
\label{a}\ee
 one finds the most general two-row Jack polynomial for $\tau_2$,
 for arbitrary $\al$,
 \bea
 \tau_2(t)
  &=&pf~m_2(t)
   \nonumber\\
&&\nonumber\\&=&
 J^{(1/\al)}_{\left(\frac{N+p-2}{2},\frac{N-p-2}{2}\right)}
 \left({t}/{\al} \right)\nonumber\\
 & &
 \nonumber\\
&=&
 c\oint\frac{dx}{2\pi i}\frac{dy}{2\pi
i}\frac{(y-x)^{2\al}}{(xy)^{\al+\frac{N}{2}}}
 e^{\sum_1^{\iy}t_i(x^i+y^i)}\left( \frac{x}{y}\right)^{p/2}
\!\!\!
\,_2F_1(\al,-p;1-\al-p;\frac{y}{x}).\nonumber\\
 & & \nonumber\\\label{b}
 \eea
 Then $\tau_{\ell}(t)$ for $\ell\geq 4$ is given by an integral of
 the same hypergeometric function in the integrand above.
%
%

\end{proposition}

\bigbreak

{\sl Acknowledgment}: We thank Michael Kleber for useful
discussions and insights.


 \section{The vector fields $\pl m / \pl t_k=\Lb^k m
+m \Lb^{\top k} $ and the finite Pfaff lattice}

The $\ell\times N$ matrix defined in (1.0.12) reads

 $$
E_{\ell,N}(t)=\left(\begin{array}{ccccc|ccc} 1&{\bf
s}_1(t)&{\bf s}_2(t)&\pp&{\bf s}_{{\ell}-1}(t)&{\bf
s}_{\ell}(t)&\pp& {\bf s}_{N-1}(t)\\ 0&1&{\bf
s}_1(t)&\pp&{\bf s}_{\ell-2}(t)&{\bf
s}_{\ell-1}(t)&\pp& {\bf s}_{N-2}(t)\\ \vdots&\vdots&
& & \vdots&\vdots& & \vdots
\\ 0&0&0&\pp&{\bf s}_1(t)&{\bf s}_2(t)&\pp
 & {\bf s}_{N-\ell +1}(t)\\
0&0&0&\pp&1&{\bf s}_1(t)&\pp
 & {\bf s}_{N-\ell }(t)
\end{array}\right)
$$

The main claim of this section can be summarized in
the following statement:

\begin{proposition} The commuting equations (for
definition of $\Lb$, see footnote 2)
\be
\frac{\pl m_{N} }{ \pl t_k}=\Lb^k m_{N} +m_{N}
\Lb^{\top k}, \label{ode}\ee
with $N\times N$ skew-symmetric initial condition
 $m(0)$, have the following solution
\be
   m_{N}(t)=E_{N,N}(t) m_N(0)
    E^{\top}_{N,N}(t).\label{momentN}
    \ee
    In particular, each $\ell \times \ell$ upper-left block of $m(t)$
    equals
\be
   m_{\ell}(t)=E_{\ell,N}(t) m_N(0)
    E^{\top}_{\ell,N}(t),\label{momentl}
    \ee

\end{proposition}

\proof Define $m_{\iy}(0)$ as the semi-infinite matrix
formed by putting $m_N(0)$ in the upper-left corner
and setting all other entries equal to 0 and let
$\Lb_{\iy}$ be the semi-infinite shift matrix. Then
the solution to the differential equations

\be \frac{\pl m_{\iy}}{\pl t_{k}}=
 \Lb_{\iy}^k m_{\iy}
+m_{\iy}\Lb_{\iy}^{\top k} \label{ode-infinity}\ee
 is given by
  \be
m_{\iy}(t)=e^{\sum^{\iy}_1t_k\Lb_{\iy}^k}
m_{\iy}(0)e^{\sum^{\iy}_1t_k\Lb_{\iy}^{\top k}}. \ee
Result (\ref{ode}) follows from the Taylor expansion
$$ e^{\sum^{\iy}_1t_k\Lb_{\iy}^k}=\sum^{\iy}_{k=0}{\bf
s}_k(t)\Lb_{\iy}^k,$$ which is an upper-triangular
semi-infinite matrix, and considering only the
upper-left $\ell\times\ell$ block.
 Each upper-left
$\ell\times\ell$ block of $m_{\iy}(t)$ for $\ell\leq
N$, equals
  \bean
m_{\ell}(t)&=&E_{\ell,\iy}(t)m_{\iy}(0)E_{\ell,\iy}^{\top}(t)\\
&=&E_{\ell,N}(t)m_{N}(0)E_{\ell,N}^{\top}(t) \eean
from which (\ref{momentl}) follows,and (\ref{momentN})
setting $\ell=N$.\qed

\remark The flow ({\ref{ode-infinity}) maintains the
finite upper-left hand corner of $m_{\iy}$ and on that
locus it is equivalent to the finite flow (\ref{ode}).
Therefore, the whole semi-infinite theory can be
applied to this case. It is possible to give a proof
of Theorem 2.1 purely within finite matrices.


\begin{theorem}
Consider the commuting equations on the $N\times N$
matrix in
\be
\frac{\pl m_N}{\pl t_i}=\Lb^im_N+m_N\Lb^i
\label{2A}\ee with skew-symmetric initial condition
$m_N(s)$ and its ``skew-Borel decomposition"
\be
m_N=Q^{-1}JQ^{-1\top}, \mbox{~for $Q\in G_{{\frak
k}}$}. \label{2B}\ee When $N$ is odd, we further
impose the differential equations for the last entry
$Q_{NN}$ of $Q$:
\be
\frac{\pl Q_{NN}}{\pl t_i}=-\frac{1}{2}Q_{N,N-i}.
\label{assumption} \ee Then for arbitrary $N>0$ the
matrix $Q$ evolves according to the equations
\be
\frac{\pl Q}{\pl t_i}Q^{-1}=-\pi_{\frak k}(Q\Lb^i
Q^{-1}) \ee and the matrix $L:=Q\Lb Q^{-1}$ provides a
solution to the Lax pair
\be
\frac{\pl L}{\pl t_i}=[-\pi_{{\frak
k}}L^i,L]=[\pi_{{\frak n}}L^i,L]. \ee
\end{theorem}

\proof For a matrix $A$, consider the projections
\bean A_0&=&\left(\begin{array}{ccccccc} \ast& &\ast&
\\
 & & & & & &O\\
\ast& &\ast& \\
 & & &\ddots\\
  & & & & &\ast&\ast\\
   &O\\
   & & & & &\ast&\ast
\end{array}
\right), \mbox{~for $N$ even}\\
&=&\left(\begin{array}{cccccccc} \ast& &\ast& \\
 & & & & & &O\\
\ast& &\ast& \\
 & & &\ddots\\
  & & & & &\ast&\ast&\\
   &O\\
   & & & & &\ast&\ast& \\
   & & & & & & &\ast
\end{array}
\right), \mbox{~for $N$ odd,} \eean and \bean
A_{00}&=&A_0, \mbox{~for $N$ even}\\
&=&\left(\begin{array}{cccccccc} \ast& &\ast& \\
 & & & & & &O\\
\ast& &\ast& \\
 & & &\ddots\\
  & & & & &\ast&\ast&\\
   &O\\
   & & & & &\ast&\ast& \\
   & & & & & & &0
\end{array}
\right), \mbox{~for $N$ odd.} \eean
The main point is to prove
that\footnote{$L_+^i:=(L^i)_+$ and $L_0^i:=(L^i)_0$.}

\bean 0&=& \frac{\pl Q}{\pl t_i}Q^{-1}+\pi_{{\frak
k}}L^i\\
 \\
&=&\frac{\pl Q}{\pl
t_i}Q^{-1}+(L^i_--J(L^i_+)^{\top}J)+\frac{1}{2}(L^i_0-J(L^i_0)^{\top}J)\\
 \\
&=:&A. \eean Also define

$$ \left(L^i+ \frac{\pl Q}{\pl
t_i}Q^{-1}\right)-J\left(L^i+ \frac{\pl Q}{\pl
t_i}Q^{-1}\right)^{\top}J=:B. $$ we have, setting
${}^. =\displaystyle{\frac{\pl}{\pl t_i}}$,

\bean 0&=& Q\left(\Lb^im+m\Lb^{\top i}-\frac{\pl
m}{\pl t_i}\right)Q^{\top}\\
 \\
&=&(Q\Lb^iQ^{-1})J+JQ^{-1\top}\Lb^{\top
i}Q^{\top}+(\dot QQ^{-1})J+JQ^{-1\top}\dot Q^{\top}\\
 \\
&=&(L^i+\dot QQ^{-1})J+J(L^i+\dot QQ^{-1})^{\top}.
\eean
Hence\footnote{$A_{-,00}=A_-+A_{00}$.}
\bean 0&=& \left(Q\left(\Lb^im+m\Lb^{\top i}-\frac{\pl
m}{\pl t_i}\right)Q^{\top}\right)_{-,00}\\
 \\
&=&\left(\left((L^i+\dot QQ^{-1})-J(L^i+\dot
QQ^{-1})^{\top}J\right)J\right)_{-,00}\\
 \\
&=&\left((L^i+\dot QQ^{-1})-J(L^i+\dot
QQ^{-1})^{\top}J\right)_{-,00}J\\ &=&B_{-,oo}J. \eean
Therefore
$$ 0=B_{-,00}J^2=\left\{\begin{array}{ll}
B_{-,0},&\mbox{ for $N$ even}\\
\\
B_{-,00}\left(\begin{array}{cc} I_{N-1}&O\\ O&0
\end{array}\right)&\mbox{for $N$ odd}
\end{array}
\right. $$ and so
\be
B_-=0\quad\mbox{and}\quad B_{00}=0. \ee
But
\bea B_-&=&(L^i+\dot QQ^{-1} -J(L^i_+)^{\top}J)_-\no\\
&=&(\dot QQ^{-1})_-+((L^i)_--J(L^i_+)^{\top}J)\no\\
&=&A_- \label{A}
 \eea
  and
   \bea B_{00}&=&2(\dot
QQ^{-1})_{00}+(L^i-J(L^i)^{\top}J)_{00}\no\\
&=&2A_{00}. \label{B}\eea
Then, by (2.0.12) and (2.0.13),
$$
0=B_-+\frac{1}{2}B_{00}=A_-+A_{00}=A_-+A_{00}+A_+,\quad\mbox{since
$A_+=0.$} $$ Therefore, when \underline{$N$ is even},
$A=0$ and the proof is finished. When \underline{$N$
is odd}, we have

$$ A=0,\mbox{~except for the $(N,N)$-entry}. $$
But the $(N,N)$th entry of $L^i$ is given by, since
$Q$ is lower-triangular,
$$
(L^i)_{NN}=(Q\Lb^iQ^{-1})_{NN}=\frac{Q_{N,N-i}}{Q_{NN}},
$$ and thus we have, using the fact that the $(N,N)$th
entry of $J(L^i)_0J$ vanishes,

\bean A_{NN}&=&\frac{\pl}{\pl t_i}\log
Q_{NN}+\frac{1}{2}(L^i)_{NN}\\
&=&\frac{1}{Q_{NN}}\left(\frac{\pl Q_{NN}}{\pl t_i}+\frac{1}{2}
Q_{N,N-i}\right),\\ &=&0, \mbox{~~by the assumption
(\ref{assumption}),} \eean thus ending the proof of Theorem
2.2.\qed


 \section{The solution to the Pfaff lattice with
   anti-diagonal skew-symmetric initial condition
 }

Consider the equations
 \be
\frac{\pl m_{N} }{ \pl t_i}=\Lb^i m_{N} +m_{N}
\Lb^{\top i},\label{ode1} \ee
  with initial
condition,
  \bean m_N(0)&=&\left(
\begin{array}{cccccc}
 &O& & & &b_{\frac{N-2}{2}}\\
 & & & &\diagup& \\
 & & & b_0& & \\
& &-b_0& & & \\
 &\diagup& & & & \\
-b_{\frac{N-2}{2}}& & & &O&
\end{array}
\right)   ~~\mbox{,  for $N$ even,} \nonumber \\ &&
 \nonumber\\
   &=&\left(
\begin{array}{ccccccc}
 &&O& & & &b_{\frac{N-3}{2}}\\
 && & & &\diagup& \\
 && & & b_0& & \\
 && &0& & & \\
&& -b_0&& & & \\
 &\diagup&& & & & \\
-b_{\frac{N-3}{2}}&& & & &O&
\end{array}
\right)
  ~~\mbox{,  for $N$ odd.}
   \eean   \vspace{-1.0cm} \be \label{initial condition}\ee

\bigbreak

 \begin{proposition} The system of equations (\ref{ode1}), with
 initial condition (\ref{initial condition}) have for solution the matrix
 $m_N(t)$, with entries, for $0\leq \ell < k \leq N$,
  \bea \lefteqn{ \mu_{\ell,k}(t)
=-\sum_{j=0}^{[\frac{N-2}{2}]-k} {\bf s}_j {\bf
s}_{N-\ell-k-j-1}(b_{[\frac{N-2}{2}]-k-j}
 -b_{[\frac{N-2}{2}]-\ell-j})  }\nonumber\\
 &&~~~~~~~~~~~
  -\sum_{[\frac{N-2}{2}]-k+1}^{[\frac{N-2}{2}]-\ell}
 {\bf
s}_j {\bf s}_{N-\ell-k-j-1}(
 -b_{[\frac{N-2}{2}]-\ell-j}). \label{mu}
 \eea
In particular
\bea \mu_{01}(t)&=&\sum_{i=0}^{\frac{N-2}{2}} b_i
 {\bf s}_{\frac{N-2}{2}+i,\frac{N-2}{2}-i}(t)
 ,~~\mbox{ for $N$ even}\nonumber\\
 &=&
  \sum_{i=0}^{\frac{N-3}{2}} b_i
 {\bf s}_{\frac{N-1}{2}+i,\frac{N-3}{2}-i}(t)
  ,~~\mbox{ for $N$ odd.}\eea

\end{proposition}

\proof Equation (\ref{mu}) is established by explicit
computation of
\begin{eqnarray*}
m_N(t)  &=&E_{N,N}(t)m_{N}(0)E_{N,N}(t)^{\top}\\
&=&\Bigl(\sum^{N-\ell-1}_{i,j=0}{\bf
s}_i(t)\mu_{i+\ell ,j+k}(0){\bf
s}_j(t)\Bigr)_{0\leq\ell,k\leq N-1}.
\end{eqnarray*}
From (\ref{mu}), one computes, for $N$ even,
 \bean \mu_{01}(t)
&=&{\bf s}_{0}{\bf s}_{N-2}
(b_{\frac{N}{2}-1}-b_{\frac{N}{2}-2}
 )
  +{\bf s}_{1}{\bf s}_{N-3} (b_{\frac{N}{2}-2}-b_{\frac{N}{2}-3})
 +  \\&&\ldots+
 {\bf s}_{\frac{N}{2}-2 }{\bf s}_{\frac{N}{2}}(b_1-b_0)+
  ({\bf s}_{\frac{N}{2}-1})^2b_0\\
  &=&
 \sum_{i=0}^{\frac{N}{2}-1} b_i
  ({\bf s}_{\frac{N}{2}-1-i}
   {\bf s}_{\frac{N}{2}-1+i}
   - {\bf s}_{\frac{N}{2}-2-i}
     {\bf s}_{\frac{N}{2}+i})  \\
 &=&  \sum_{i=0}^{\frac{N}{2}-1} b_i
 {\bf s}_{\frac{N}{2}-1+i,\frac{N}{2}-1-i}(t),
 \eean
and for $N$ odd,
 \bean \mu_{01}(t)
&=&{\bf s}_{0}{\bf s}_{N-2}
(b_{\frac{N-3}{2}}-b_{\frac{N-5}{2}}
 )
  +{\bf s}_{1}{\bf s}_{N-3} (b_{\frac{N-5}{2}}-
  b_{\frac{N-7}{2}})
 +   \\&& \ldots+
 {\bf s}_{\frac{N-5}{2} }{\bf s}_{\frac{N+1}{2}}(b_1-b_0)+
  {\bf s}_{\frac{N-3}{2}}  {\bf s}_{\frac{N-1}{2}}b_0\\
  &=&
  \sum_{i=0}^{\frac{N-3}{2}} b_i
 {\bf s}_{\frac{N-1}{2}+i,\frac{N-3}{2}-i}(t),
\eean
 ending the proof of Proposition 3.1.\qed
 Define\footnote{$\vr_{i,j}$ denotes the
 matrix with all
 zero entries, except for a $1$ at the $(i,j)$th entry.}
 \bean
 m_N(0;z)&:=& m_N(0)
 ~~,~~~~\mbox{ for $N$ even,} 
\\
m_N(0;z)&:=&m_N(0) +z^2
\vr_{\frac{N+1}{2},\frac{N+1}{2}} ~~,~~~~~\mbox{ for
$N$ odd,}\\
 &=&\left(
\begin{array}{ccccccc}
 &&O& & & &b_{\frac{N-3}{2}}\\
 && & & &\diagup& \\
 && & & b_0& & \\
 && &z^2& & & \\
&& -b_0&& & & \\
 &\diagup&& & & & \\
-b_{\frac{N-3}{2}}&& & & &O&
\end{array}
\right)\\
   \eean   \vspace{-1.4cm} \be \ee
\begin{proposition}

\bean \lefteqn{ \det {}^{1/2}\left( E_{\ell,N}(t)
m_N(0;z)
    E^{\top}_{\ell,N}(t)\right)}\\
    &=&
     z^{\eta (N, \ell)}\sum_{\lb\in\BY^{(\ell)}_{\frac{\ell
(N-\ell)}{2}}}\left(\prod_1^{[\ell/2]}b_{\lb_
i-i+\ell-\left[\frac{N+1}{2}\right]} \right){\bf
s}_{\lb_1 \geq \ldots \geq \lb_{\ell}}(t).
 \eean
 with
 $$
 \eta(N,\ell)=\left\{
 \begin{array}{l}
 1~~, ~\mbox{for $N$ and $\ell$ odd.}\\
 0~~, ~\mbox{otherwise}.
 \end{array}\right.
 $$
\end{proposition}

\bigbreak

\begin{lemma}

Consider an arbitrary $N\times N$ matrix
$A=(A_{ij})_{1\leq i,j\leq N}$, with
$r=\displaystyle{\left[\frac{N}{2}\right]}$ and
$A_{\ell}:=(A_{ij})_{1\leq i\leq\ell\atop{1\leq j\leq
N}}$ and consider the anti-diagonal matrix

\bean m_N&=&\left(\begin{array}{ccccccc}
 & & & & &c_r& \\
 & & & &\diagup\\
 &O\\
 & & &c_1\\
 & &-c_1\\
 &\diagup & & &O\\
-c_r
\end{array}\right) \mbox{~for $N$ even}\\
 \\
&=&\left(\begin{array}{cccccccc}
 & & & & & &c_r& \\
 & & & & &\diagup\\
 &&O\\
 && & &c_1\\
 && & z^2\\
 && -c_1\\
 &\diagup & & &O\\
-c_r &&&&&
\end{array}\right)\mbox{~for $N$ odd.}
\eean
Setting\footnote{$B_{(j_1,\ldots,j_n)}$ denotes the
matrix formed with the columns $j_1,\dots,j_n$ of $B$}

$$ m^A_{\ell}(z):=A_{\ell}m_N(z)A_{\ell}^{\top} $$ and

\newpage


\bea
 \lefteqn{P_{N,\ell}=\sum_{1\leq
i_1\leq\ldots\leq i_{[\ell/2]}\leq r}c_{i_1}\ldots
c_{i_{[\ell/2]}}}\no\\ &&\times
\left\{\begin{array}{l} \det(A_{\ell})
 _{(r-i_{[\ell/2]}+1,\ldots,r-i_1+1,r+i_1,
\ldots,r+i_{[\ell/2]})}
~~\mbox{~for $N$ even, $\ell$ even}\no\\  \\
  \det(A_{\ell})
 _{(r-i_{[\ell/2]}+1,\ldots,r-i_1+1,r+i_1+1,
 \ldots,r+i_{[\ell/2]}+1)}
 ~~\mbox{~for $N$ odd, $\ell$ even}\\  \\
\\ \det(A_{\ell})
 _{(r-i_{[\ell/2]}+1,\ldots,r-i_1+1,r+1,r+i_1+1,\ldots,r+i_{[\ell/2]}
+1)}~~\mbox{~for $N$ odd, $\ell$ odd}\\
\end{array}
\right.\no\\ \label{Bonnet}\eea
 we have

$$ \det m_{\ell}^A=\left\{\begin{array}{cl}
0&\mbox{for $N$ even, $\ell$ odd}\\
\\
(pf~m_{\ell}^A)^2=(P_{N,\ell})^2&\mbox{for $N$ even,
$\ell$ even}\\
\\
z^2P^2_{N,\ell}&\mbox{for $N$ odd, $\ell$ odd}\\
\\
(pf~m_{\ell}^A(0))^2=(P_{N,\ell})^2&\mbox{for $N$ odd,
$\ell$ even.}
\end{array}
\right. $$
\end{lemma}

\proof Let $w_i\in\BC^{\ell}$ be the columns of
$A_{\ell}$

$$ A_{\ell}=[w_0,w_1,...,w_{2r}], $$ and observe

\bean
m_{\ell}^A(z)=A_{\ell}m_N(z)A_{\ell}^{\top}
&=&A_{\ell}\left(z^2\vr_{
r+1,r+1}+m_N(0)\right)A_{\ell}^{\top}\\
&=&z^2w_r\otimes w_r+m_{\ell}^A(0). \eean Let $U$ be a
$\ell\times\ell$ matrix, rational in the $a_{ij}$,
such that

$$ U ~w_r=\al e_1,\quad\det U=1. $$ Then, using
$U(x\otimes y)V=(Ux)\otimes(V^{\top}y)$ and setting
$M:=U ~m_{\ell}^A(0)U^{\top}$, which is
skew-symmetric, we find

\bean \det m_{\ell}^A(z)&=&\det
U~m_{\ell}^A(z)U^{\top}\\ &=&\det\left(z^2U(w_r\otimes
w_r)U^{\top}+U~m_{\ell}^A(0)U^{\top}\right)\\
&=&\det\left(z^2\al^2e_1\otimes
e_1+U~m_{\ell}^A(0)U^{\top}\right),\\
&=&\det\left(\begin{array}{c|cccc}
(z\al)^2&M_{12}&M_{13}&...&M_{1\ell}\\ \hline
-M_{12}&0&M_{23}&...&M_{2\ell}\\ -M_{13}&-M_{23}&  \\
\vdots&\vdots& & &\vdots\\ -M_{1\ell}&-M_{2\ell}&...&
&0
\\
\end{array}
\right)\\ &=&(z\al)^2\det (M_{ij})_{2\leq
i,j\leq\ell}+\det(M_{ij})_{1\leq i,j\leq\ell}, \eean
with $M_{ij}=-M_{ji}$. Therefore

\be
\begin{array}{lll}
\det m_{\ell}^A(z)&=\det
m_{\ell}^A(0)=(pf~m^A_{\ell}(0))^2,&\mbox{~for $\ell$
even}\\
 \\
&=(z\al)^2\det(M_{ij})_{2\leq i,j\leq\ell} = (z \al
~pf(M_{ij})_{2\leq i,j\leq \ell})^2, &\mbox{~for
$\ell$ odd,}
\end{array}
\label{expression}\ee
  the latter being the square of a polynomial in
$z$, the $c_i$ and the entries of the matrix $A$.
\newpage

Using the Cauchy-Bonnet formula twice, one computes,
say, for $N$ and $\ell$ odd,

\medbreak\noindent $\det m_{\ell}^A(z)=\det
A_{\ell}m_N(z)A_{\ell}^{\top}$ \bean &=&\sum_{1\leq
\al_1<...<\al_{\ell}\leq
N}\det\left((A_{\ell})_{i,\al_j}\right)_{1\leq
i,j\leq\ell}\det\left((A_{\ell}m^{\top})_{i,\al_j}\right)_{1\leq
i,j\leq\ell}\\ & &\\ &=&
\sum_{1\leq\al_1<...<\al_{\ell}\leq
N\atop{1\leq\beta_1<...<\beta_{\ell}\leq
N\atop{\al_i+\beta_{\ell-i+1}=N+1}
 }}\det\left((A_{\ell})_{i,\al_j}\right)_{1\leq
i,j\leq\ell}\det\left((A_{\ell})_{i,\beta_j}\right)_{1\leq
i,j\leq\ell}\det\left((m^{\top})_{\beta_i,\al_j}
 \right)_{1\leq i,j\leq\ell}\\ & &\\
&=&\sum_{1\leq\al_1<...<\al_{\ell}\leq
N\atop{1\leq\beta_1<...<\beta_{\ell}\leq
N\atop{\al_i+\beta_{\ell-i+1}=N+1\atop{{\rm for~}1\leq
i\leq\ell}}}}
\det\left((A_{\ell})_{i,\al_j}\right)_{1\leq
i,j\leq\ell}\det\left((A_{\ell})_{i,\beta_j}\right)_{1\leq
i,j\leq\ell}\det(m_{\al_i,\beta_j})_{1\leq
i,j\leq\ell}\\ & & \\
&=&\left(\sum_{1\leq\al_1<...<\al_{\ell}\leq
N\atop{1\leq\beta_1<...<\beta_{\ell}\leq
N}}+\sum_{1\leq\al_1<...<\al_{\ell}\leq
N\atop{1\leq\beta_1<...<\beta_{\ell}\leq
N}}\right)\det\left((A_{\ell})_{i,\al_j}\right)_{1\leq
i,j\leq\ell}\det\left((A_{\ell})_{i,\beta_j}\right)_{1\leq
i,j\leq\ell}\\ & &\phantom{
}_{(\al_1,...,\al_{\ell})=(\beta_1,...,\beta_{\ell})
\atop{\al_i+\beta_{\ell -i+1}=N+1\atop{{\rm for~}1\leq
i\leq\ell}}}~ \phantom{ }_{(\al_1,...,\al_{\ell})\neq
(\beta_1,...,\beta_{\ell}) \atop{\al_i+\beta_{\ell
-i+1}=N+1\atop{{\rm for~}1\leq i\leq\ell}}}\\ &
&\hspace*{9cm}\times \det(m_{\al_i,\beta_j})_{1\leq
i,j\leq\ell}\\
 &\stackrel{*}{=}&
  z^2\sum_{1\leq\al_1
<...<\al_{\frac{\ell
-1}{2}}<\frac{N+1}{2}\atop{\al_{\frac{\ell
+1}{2}+i}+\al_{\frac{\ell +1}{2}-i}=N+1\atop{{\rm
for~}0\leq i\leq\frac{\ell
-1}{2}}}}c^2_{\frac{N+1}{2}-\al_1
}...c^2_{\frac{N+1}{2}-\al_{\frac{\ell -1}{2}}}
 {\rm det}^2\left((A_{\ell})_{i,\al_j}\right)_{1\leq i,j\leq\ell}+...\\
 &=&
  \left(z\sum_{1\leq\al_1 <...<\al_{\frac{\ell
-1}{2}}<\frac{N+1}{2}\atop{ \al_{\frac{\ell
+1}{2}+i}+\al_{\frac{\ell+1}{2}-i}=N+1}}c_{\frac{N+1}{2}-\al_1}...
c_{\frac{N+1}{2}-\al_{\frac{\ell
-1}{2}}}\det\left((A_{\ell})_{i,\al_j}\right)_{1\leq
i,j\leq\ell}\right)^2
 \mbox{using (\ref{expression})}  \\
  &=&\left(z\sum_{1\leq i_1<...<
i_{\frac{\ell
-1}{2}}\leq\frac{N-1}{2}}c_{i_{\frac{\ell-1}{2}}}...
c_{i_1}\det(A_{\ell})_{\left(\frac{N+1}{2}-i_{\frac{\ell-1}{2}},...,\frac{N+1}{2
}-i_1,
\frac{N+1}{2},\frac{N+1}{2}+i_1,...,\frac{N+1}{2}+i_{\frac{\ell-1}{2}}
 \right)}\right)^2. \eean

\newpage

In $\stackrel{*}{=}$ we have used the fact that

\be \left.
\begin{array}{c}
(\al_1,...,\al_{\ell})=(\beta_1,...,\beta_{\ell})\\
\al_i+\beta_{\ell-i+1}=N+1
\end{array}\right\}\DF\left\{\begin{array}{c}
\al_{\frac{\ell +1}{2}+i}+\al_{\frac{\ell
+1}{2}-i}=N+1,\\ \mbox{for~} 0\leq i\leq\frac{\ell
-1}{2}\\ \beta_{\ell -i+1}=N+1-\al_i.
\end{array}\right.
\label{1}\ee


Indeed, for \underline{$N$ odd}
 consider sequences
$\al_i$ symmetric about
\be
\al_{\frac{\ell+1}{2}}=\frac{N+1}{2} \label{2}\ee
  i.e.
\be
\al_{\frac{\ell+1}{2}+i}+\al_{\frac{\ell+1}{2}-i}=
N+1
,~~\mbox{for}~~0\leq i\leq \frac{\ell-1}{2}.
\label{3}\ee
Then, using (\ref{1}) and (\ref{3})
 $$
  \beta_{\frac{\ell
+1}{2}-i}=N+1-\al_{\frac{\ell
+1}{2}+i}
=\al_{\frac{\ell +1}{2}-i}, $$ thus implying $$
(\al_1,...,\al_{\ell})=(\beta_1,...,\beta_{\ell}). $$
Vice versa, the latter implies (\ref{1}) and thus
(\ref{2}). This establishes Lemma 3.3 for the case $N$
and $\ell$ odd; for the other cases, one proceeds in a
similar fashion.


\bigbreak

\noindent {\it Proof of Proposition 3.2:} ~Apply Lemma 3.3 to
$A_{\ell}=E_{\ell,N}(t)=(\gs_{j-i})_{{1\leq i \leq \ell}\atop
{1\leq j \leq N}}$, with $1\leq k_1<k_2<\ldots < k_{\ell}$:

\bea
\det(A_{\ell})_{k_1,...,k_{\ell}}&=&\det\left(\begin{array}{cccc}
\gs_{k_1-1}&...&\gs_{k_{\ell-1}-1}&\gs_{k_{\ell}-1}\\
\vdots& &\vdots&\vdots\\
\gs_{k_1-\ell}&...&\gs_{k_{\ell-1}-\ell}&\gs_{k_{\ell}-\ell}
\end{array}\right)\nonumber\\
& & \nonumber\\ &=&\det\left(\begin{array}{cccc}
\gs_{k_{\ell}-\ell}&\gs_{k_{\ell}-\ell+1}&...&\gs_{k_{\ell}-1}\\
\gs_{k_{\ell-1}-\ell}&\gs_{k_{\ell-1}-\ell+1}&...&\gs_{k_{\ell-1}-1}\\
\vdots& & &\vdots\\ \gs_{k_1-\ell}& &...&\gs_{k_1-1}
\end{array}\right)\nonumber\\
&=&\gs_{k_{\ell}-\ell,k_{\ell-1}-\ell+1,...,k_1-\ell+(\ell-1)}\nonumber\\
&=&\gs_{\lb_1\geq ...\geq \lb_{\ell}}\nonumber\\
&=&\gs_{\lb} \label{4}\eea
where
\be \lb_i=k_{\ell-i+1}-\ell+i-1,\mbox{~for~}1\leq
i\leq\ell. \label{5}\ee

In order to apply Lemma 3.3, the $k_i$ inherent in formula
(\ref{Bonnet}) must be as in formula (\ref{4}); i.e., setting
$r=[N/2]$, the $k_j$'s must satisfy
  \be
  k_j=\left[ \frac{N}{2}\right]- i_{[\ell/2]-j+1}+1=
   N+1-k_{\ell-j+1}  , ~~\mbox{for}~1\leq j\leq
   \left[\frac{\ell+1}{2}\right] \label{6}
   \ee
and thus
 \bean
 i_{[\ell/2]+1-j}-1&=& k_{\ell+1-j}-\left[\frac{N+1}{2}\right]-1
\\
&=& \lb_j+\ell-j-\left[\frac{N+1}{2}\right]. \eean
Therefore, formula (\ref{Bonnet}) can be applied with
$$
 c_{i_{[\ell/2]-j+1}}= b_{\lb_j+\ell -j-[(N+1)/2]},
 ~~~\mbox{for}~~ 1\leq j\leq
 \left[\frac{\ell}{2}\right].
 $$
From (\ref{5}) and (\ref{6}), it follows that
 $$
\lb_i+\lb_{\ell+1-i}=k_{\ell+1-i}+k_i-\ell-1=N+1-\ell-1=N-\ell
,$$ showing that $$
 \lb \in
  \BY^{(\ell)}_{\frac{\ell
(N-\ell)}{2}} ,$$
 establishing Proposition 3.2. \qed


 \section{Proof of Theorem 1.1}

Using the standard notation for the partition
$1^j=\overbrace{(1,\ldots,1)}^j$, we state
\begin{lemma}

For
\be
\left.\begin{array}{c} {\bf s}_i(-\tilde\pl){\bf
s}_{1^j}(t)\\
 \\
 \left(-\frac{\pl}{\pl t_i}\right){\bf s}_{1^j}(t)
\end{array}
\right\}=(-1)^i{\bf s}_{1^{j-i}}(t). \label{Schur}\ee
\end{lemma}

\proof Using the usual inner-product between symmetric
functions, we have

\bean {\bf s}_i(\tp)\gs_j(t)&=&\la \gs_i(t+u)\cdot
1,\gs_j(t+u)\ra \\ &=&\la\gs_j(t+u),\gs_i(t+u)\cdot
1\ra \\ &=&\la \gs_{j-i}(t+u),1\ra \\ &=&\la
1,\gs_{j-i}(t+u)\ra \\ &=&\gs_{j-i}(t+u)\big|_{u=0}\\
&=&\gs_{j-i}(t) \eean and so, changing $t\mapsto -t$,
$$ \gs_i(-\tp)\gs_j(-t)=\gs_{j-i}(-t), $$ from which
this first relation follows upon noticing that
 \be
\gs_j(-t)=(-1)^j\gs_{1^j}(t).\label{7}
  \ee
  This last relation (\ref{7})
also leads to the second identity (\ref{Schur}), using
$(\pl /\pl t_i)\gs_j(t)=\gs_{j-i}(t)$.\qed

\bigbreak

 \noindent {\it Proof of Theorem 1.1:}~By Proposition
2.1, the equations for the $N\times N$ matrix $m_N$
$$
  \frac{\pl m_N}{\pl t_k}=\Lb^km_N+m_N\Lb^k,
   $$
    with
skew-symmetric initial condition $m_N(0)$ has the
following solution $$
m_N(t)=E_{\ell,N}m_N(0)E^{\top}_{\ell,N}(t), $$ which
remains skew-symmetric in time. Define a $t$-dependent
skew-inner product such that $\la
y^i,z^j\ra_t=m_{ij}(t)$,
i.e.\footnote{$\chi(y):=(1,y,y^2,\ldots)^{\top}.$}, $$
\la\chi_N(y)\chi(z)^{\top}\ra =m_N(t). $$ Performing
the skew Borel decomposition

\be
m_N(t)=Q^{-1}(t)J Q^{-1\top},\quad\mbox{with~}Q(t)\in
G_k \ee is tantamount to the process of finding a
finite set of skew-orthonormal polynomials; that is,
satisfying $$ \Big(\la
q_i(t;z),q_j(t;z)\ra\Big)_{1\leq i,j\leq N}=J. $$
Indeed, the polynomials $q_i(t;z)$ in $z$, depending
on $t$, $$ \left(\begin{array}{c} q_0\\ q_1\\ \vdots\\
q_{N-1}
\end{array}
\right)=Q\left(\begin{array}{c} 1\\ z\\ \vdots\\
z^{N-1}
\end{array}
\right) $$ satisfy \bean \Big(\la
q_i(t;y),q_i(t;z)\ra\Big)_{0\leq i,j\leq N-1}&=&\la
Q(t)\chi_N(y),Q(t)\chi_N(z)\ra
\\
&=&\la Q(t)\chi_N(y)\chi_N(z)Q^{\top}(t)\ra \\
&=&Q(t)\la\chi_N(y)\chi_N(z)\ra Q^{\top}(t)\\
&=&Q(t)m_N(t)Q^{\top}(t)\\ &=&J. \eean According to
\cite{AvM6}, the skew-orthogonal polynomials are
related to the $\tau$-functions $(\tau_0=1$,
$\tau_N=c$) $$ \tau_{\ell}(t)=pf~m_{\ell}(t) $$ as
follows \bean
q_{2n}&=&\frac{z^{2n}}{\sqrt{\tau_{2n}\tau_{2n+2}}}\tau_{2n}(t-[z^{-1}])\\
q_{2n+1}&=&\frac{z^{2n}}{\sqrt{\tau_{2n}\tau_{2n+2}}}\left(z+\frac{\pl}{
\pl t_1}\right)\tau_{2n}(t-[z^{-1}]),~~0\leq 2n\leq N-2.
  \eean
This ends the proof of Theorem 1.1 for $N$ even. However for odd
$N$, we must verify condition (2.0.8) of Theorem 2.2. This
requires knowing $q_{N-1}(t;z)$ explicitly. For later purposes we
shall also need $q_{N-1}(t;z)$ for even $N$.

\bigbreak

\underline{For $N$ even}, $q_{N-1}$ takes on the
following form $$
q_{N-1}(t;z)=\frac{z^{N-2}}{\sqrt{\tau_{N-2}\tau_N}}\left(z+\frac{\pl}{
\pl t_1}\right)\tau_{N-2}(t-[z^{-1}]), $$
 with (using Proposition 3.2)
$$
\tau_{N-2}(t)=\sum_{\lb\in\BY_{N-2}^{(N-2)}}
 \left(\prod_1^{\frac{N-2}{2}}
b_{\lb_i-i+\frac{N}{2}-2}\right)\gs_{\lb}(t),
 $$
  where
 $$
\BY_{N-2}^{(N-2)}=\left\{1^{N-2},(2,1^{N-4}),(2^2,1^{N-6})
,...,(2^i,1^{N-2i-2}),... \right\}. $$

\vspace*{1cm}

\underline{For $N$ odd}, $q_{N-1}$ has the form

$$
q_{N-1}(t;z)=\frac{z^{N-1}}{\sqrt{\tau_{N-1}}}\tau_{N-1}(t-[z^{-1}])
$$ with
 \be
\tau_{N-1}(t)=b_0...b_{\frac{N-3}{2}}\gs_{\left(1^{\frac{N-1}{2}}\right)}(t).
 \ee
 Indeed, observe that the set of partitions
 \bean
\BY^{\ell}_{\frac{\ell(N-\ell)}{2}}\Biggl|_{\ell=N-1}=\BY^{(N-1)}_{\frac{N-1}{2}
} &=&\left\{\begin{array}{l}
(\lb_1,...,\lb_{N-1})\in\BY_{\frac{N-1}{2}}\\
 \\
\mbox{with~}\lb_i+\lb_{\ell+1-i}=1
\end{array}\right\}\\
 \\
&=&\left\{1^{\frac{N-1}{2}}\right\} \eean consists of
one element $1^{\frac{N-1}{2}}$. Therefore, setting
$\lb_i=1$ for $1\leq i\leq \frac{N-1}{2}$ one finds,
again by Proposition 3.2,
$$
\tau_{N-1}(t)=b_0...b_{\frac{N-3}{2}}\gs_{\left(1^{\frac{N-1}{2}}\right)}(t).
$$
 The last row of $\tilde Q$ is given by:
%
\bean
 \sum_0^{N-1}\tilde Q_{N,j+1}z^j&=&
z^{N-1}\tau_{N-1}(t-[z^{-1}])\\
 &=&\sum_{i=0}^{N-1}\gs_i(-\tp)\tau_{N-1}(t)z^{N-1-i}\\
& & \\
&=&b_0...b_{\frac{N-3}{2}}\sum_{i=0}^{N-1}\gs_i(-\tp)\gs_{\left(1^{\frac{N-1}{2}
}\right)}(t)z^{N-1-i}\\ & & \\
&=&b_0...b_{\frac{N-3}{2}}\sum_{i=0}^{\frac{N-1}{2}}z^{N-1-i}(-1)^i
\gs_{\left(1^{\frac{N-1}{2}-i}\right)}(t),
 \eean
  using Lemma 4.1, and so
  $$ \tilde
Q_{N,N-i}=(-1)^i\left(\prod_0^{\frac{N-3}{2}}b_k\right)\gs_{\left(1^{\frac{N-1}{
2}-i}\right)}.
  $$
 So, the last row of $\tilde Q$
reads
$$
\prod_0^{\frac{N-3}{2}}b_i\left(\underbrace{0,...,0}_{\frac{N-1}{2}},(-1)^{\frac{N-1}{2}},(-1)^{\frac{N-3}{2}}\gs_1(t),(-1)^{\frac{N-5}{2}
}\gs_{(1^2)}(t),...,
\gs_{\left(1^{\frac{N-1}{2}}\right)}(t)\right) $$ and
and the last row of $Q=D\tilde Q$:
  \bean
  Q_{N,N-i}=(D\tilde
Q)_{N,N-i}&=&(-1)^i\prod_0^{\frac{N-3}{2}}b_k
\frac{\gs_{\left(1^{\frac{N-1}{2}-i}\right)}(t)}{\sqrt{\tau_{N-1}}}\\
& & \\
&=&(-1)^i\left(\prod_0^{N-3}b_k\right)^{1/2}\frac{\gs_{\left(1^{\frac{N-1}{2}-i}
\right)}(t)}
{\left(\gs_{\left(1^{\frac{N-1}{2}}\right)}(t)\right)^{1/2}}
  \eean
  and so, using Lemma 4.1,
$$
\frac{\pl Q_{N,N}}{\pl t_i}=-\frac{(-1)^i}{2}
\left(\prod_0^{N-3}b_k\right)^{1/2}
\frac{\gs_{\left(1^{\frac{N-1}{2}-i} \right)}(t)}
{\left(\gs_{\left(1^{\frac{N-1}{2}}\right)}(t)
\right)^{1/2}}= -\frac{1}{2}Q_{N,N-i}.
 $$
  Having checked (\ref{2A}), (\ref{2B}) and
  (\ref{assumption}) (in the odd case) of Theorem 2.2, we have found a
  solution of the Pfaff lattice. This finally concludes the proof
  of Theorem 1.1.\qed


\medskip\noindent{\it Proof of Theorem 1.2:\/}
 According to \cite{AvM6}, Pfaff $\tau$-functions satisfy
bilinear relations\footnote{
    $\tilde\pl=(\pl/\pl t_1,(1/2)\pl/\pl t_2,(1/3)\pl/\pl t_3,\dots)$,
    $\tilde D=\left(D_1,(1/2)D_2,(1/3)D_3,\dots\right)$ is the
    corresponding Hirota symbol: $P(\tilde D)f\cdot g :=
    P(\pl/\pl y_1,(1/2)\pl/\pl y_2,\dots)f(t+y)g(t-y)|_{y=0}$,
    and ${\bf s}_k$ are the previously defined elementary Schur functions:
    $\sum_{k=0}^\infty {\bf s}_k(t)z^k:=
     \exp(\sum_{i=1}^\infty t_iz^i)$.
 For further notations, see Dickey \cite{D}.}: for all $t,t'\in \BC^{\iy}$ and $m,n$ positive
integers
\begin{multline}
\oint_{z=\iy}\tau_{2n}(t-[z^{-1}])\tau_{2m+2}(t'+[z^{-1}])
e^{\sum_{i=0}^\iy(t_i-t'_i)z^i} z^{2n-2m-2}dz
\\
{}+\oint_{z=0}\tau_{2n+2}(t+[z])\tau_{2m}(t'-[z])
e^{\sum_{i=0}^\iy(t'_i-t_i)z^{-i}}z^{2n-2m}dz=0\,,\nonumber
\end{multline}
Shifting appropriately and taking residues leads to
the ``differential Fay identity'':
\begin{multline}
\{\tau_{2n}(t-[u]),\tau_{2n}(t-[v])\}
\\
{}+(u^{-1}-v^{-1})(\tau_{2n}(t-[u])\tau_{2n}(t-[v])
-\tau_{2n}(t)\tau_{2n}(t-[u]-[v]))
\\
=uv(u-v)\tau_{2n-2}(t-[u]-[v])\tau_{2n+2}(t),
\label{3.3}
\end{multline}
and Hirota bilinear equations, involving nearest
neighbors:
\begin{equation}
\left({\bf s}_{k+4}(\tilde
 \pl)-\frac{1}{2}\frac{\pl}{\pl t_1}\frac{\pl}{\pl
t_{k+3}} \right)\tau_{2n}\cdot \tau_{2n}={\bf
 s}_k(\tilde
 \pl) \tau_{2n+2}\cdot \tau_{2n-2}. 
 \end{equation}
It only remains to check the``boundary condition":
   \be
 \left\{\begin{array}{ll}
\displaystyle{\tau_N=\prod_0^{\frac{N-2}{2}} b_i},
 &\mbox{for even $N$,}\\ \\ \tau_{N+1}=0
 ,&\mbox{for odd $N$.}
\end{array} \right.
 \ee
 Indeed, for even $N$,
 using $\det
    E_{NN}(t)=1$ and the matrix (3.0.2), we have that
$$
 (pf~ m_N(t))^2=\det \left( E_{N,N}(t) m_N(0)
    E^{\top}_{N,N}(t)\right)=\det m_N(0)=
    \prod_0^{\frac{N-2}{2}} b_i.  $$
Moreover, for odd $N$, according to (4.0.4), $\tau_{N-1}$ is a
pure Schur polynomial, which is known to satisfy the KP Fay
identity; i.e., the equation (4.0.5), without right hand side.
This justifies setting $\tau_{N+1}=0$ for odd $N$. \qed


In the next Proposition, we show that the finite vector of
skew-orthogonal polynomials form an eigenvector of the matrix $L$,
with modified boundary condition.

\begin{proposition} For even $N$, the skew-orthonormal polynomials
$q=(q_0,.\\..,q_{N-1})^{\top}=Q\quad (1,...,z^{N-1})^{\top}$ are
eigenfunctions for $L$, with the boundary condition: $$
Lq=zq-(0,...,0,z^{N}){\sqrt{pf~m_{N-2}}} \left(
 {\displaystyle{\prod_0^{\frac{N-2}{2}}}b_i}
  \right)^{-1/2}.
$$
\end{proposition}

\proof
 Indeed
  \bean
   Lq&=&Q~\Lambda~Q^{-1}Q\MAT{1} 1\\
\vdots\\ z^{N-1} \mat\\ &=&Q~\Lambda\MAT{1}1\\
\vdots\\ z^{N-1}\mat\\ &=&Q~z\MAT{1}1\\ \vdots\\
z^{N-2}\\ 0\mat\\ &=&z\MAT{1}q_0\\ q_1\\ \vdots\\
q_{N-2}\\ \bar q_{N-1} \mat =zq+z(0,...,0,\bar q_{N-1}-q_{N-1}) ,
\eean
  where $\bar q_{N-1}$ is the
  same as $q_{N-1}$, but without leading term, i.e.,
  $\bar q_{N-1}=q_{N-1}-Q_{NN}z^{N-1}$, where by (4.0.7) we have
$$ Q_{NN}= \sqrt{\frac{\tau_{N-2}}{\tau_{N}}}
 =
   {\sqrt{pf~m_{N-2}
}} \left({\displaystyle{\prod_0^{\frac{N-2}{2}}}b_i}
\right)^{-1/2},$$ ending the proof of Proposition 4.2.\qed

\section{Vertex operators}


The purpose of this section is to prove Theorem 1.3
and Corollary 1.4. Define as in (\ref{beta}),
 \be
  \beta :=\frac{N}{2}-\ell +1.\label{samebeta}
 \ee
Remembering from (\ref{vertex}) the vertex operator $X^{}(t;z)$,
consider now its formal expansion in powers of $z$ \be
X^{}(t;z)=e^{\sum_{1}^\iy t_iz^i} e^{-
\sum_{1}^\iy\frac{z^{-i}}{i}\frac{\pl}{\pl t_i}} =:
 \sum_{i \in \BZ} B_i z^i
  ,
\label{vertex1} \ee
 with differential operators (see footnote 10)
  \be
  B_i: =B^{(\al)}_i\Big|_{\al=1}~~ \mbox{and}~~
  B^{(\al)}_i:=\sum_{j\geq 0} {\bf s}_{i+j}(\al t) {\bf s}_j(-\al
  \tilde \pl_{t}).
  \label{B-operators}\ee
    Also define as in (\ref{vectorvertex})
 the vector vertex operator\footnote{$\chi(z):=(z^i)_{i\geq 0}$.}
%
%
\be
\BX(t;z)=\Lambda^{\top}e^{\sum^{\iy}_1t_iz^i}e^{-\sum^{\iy}_1\frac{z^{-i}}{i}
\frac{\pl}{\pl t_i}}\chi(z).\label{vectorvertex1} \ee
 Also remember the definitions of
the {\em integrated vertex operator}, in terms of the vertex
operator (\ref{vertex1}) and a function $\rho_b$, defined in
(\ref{rho1}) below,
 \bean  Y_{\beta}(t)&:=
&
 \oiint X(t;y)X(t;z)\frac{\rho_b(y/
z)dy~dz}{z^2(yz)^{\beta}}
  \eean
and the {\em integrated vector vertex operator}, in
terms of (\ref{vectorvertex1}),
\be
\BY_N(t)  = 
 \oiint
\BX(t;y)\BX(t;z)
 \frac{\rho_b(y/
z)dy~dz}{2(yz)^{N/2}z}. \ee
 In both cases, the double
integral around two contours about $\iy$ amounts to computing the
coefficient of $1/yz$. The next Theorem is nothing but a
rephrasing of Theorem 1.3 and Corollary 1.4.


\begin{theorem} For a given set of $b_i$, the sequence of
$\tau$-functions $\tau_0,\tau_2,\tau_4,\dots$, defined in
(\ref{tau-functions}), is generated by the vertex operators
$Y_{\beta}$:
 \be
 Y_{\beta}\tau_{\ell
-2}=\ell\tau_{\ell}.
 \label{statementA}
 \ee
 The vector
$ I=(I_0,I_2,I_4,\dots),\mbox{~with
$I_{\ell}=\left(\frac{\ell}{2}\right)!\tau_{\ell}$} $ is a fixed
point for the vector vertex operator $\BY_N$, namely
 \be (\BY_N I)_{\ell}=I_{\ell},\mbox{~for even
 $\ell$.}\label{statementB}
 \ee
\end{theorem}

 We shall first need a few propositions.

\begin{proposition} Defining

\be \rho_b(x):=\left\{\begin{array}{l}
\rho_b^{(e)}(x):=\displaystyle{\sum_{i\geq
0}}b_i(x^{-i-1}-x^i),\mbox{~for $N$ even,}\\
 \\
\rho_b^{(o)}(x):=x^{-1/2}\displaystyle{\sum_{i\geq
0}}b_i(x^{-i-1}-x^{i+1}),\mbox{~for $N$ odd,}
\end{array}\right.
\label{rho1}\ee
 we have

\bea
Y_{\beta}(t)&=&\oiint
X(t;y)X(t;z)\frac{\rho_b(y/z)dy~dz}{z^2(yz)^{\beta}}\no\\
 & & \no\\
&=&\left\{\begin{array}{l} \displaystyle{\sum_{j\geq
0}}b_j(B_{\beta+j}B_{\beta-j}-B_{\beta-j-1}
B_{\beta+j+1})~,\mbox{~for $N$ even}\\
 \\
\displaystyle{\sum_{j\geq
0}}b_j(B_{\beta+j+1/2}B_{\beta-j-1/2}-B_{\beta-j-3/2}
B_{\beta+j+3/2})~, \mbox{~for $N$ odd.}
\end{array}\right.
 \label{Y-operator}
\eea
\end{proposition}

\proof Compute \underline{for $N$ even},
\bean
\frac{X(t;y)X(t;z)}{(yz)^{\beta}}&=&\sum_{i\in\BZ}B_iy^{i-\beta}\sum_{j\in\BZ}B_
jz^{j-\beta}\\ &=&\sum_{i\in\BZ}B_{\beta
+i}y^i\cdot\sum_{j\in\BZ}B_{\beta -j}z^{-j}\\
&=&\sum_{i,j\in\BZ}B_{\beta +i}B_{\beta
-j}\frac{y^i}{z^j}\\ &=&\sum_{j\in\BZ}B_{\beta
+j}B_{\beta -j}\left(\frac{y}{z}\right)^j+\sum_{i\neq
j\in\BZ}a_{ij}\frac{y^i}{z^j} \eean
  and so,
 \bean
  \lefteqn{\rho_b^{(e)}\left(\frac{y}{z}\right)\cdot
\frac{X(t;y)X(t;z)}{z^2(yz)^{\beta}}}\\
&=&\frac{1}{z^2}\left(\sum_{i\geq
0}b_i\left[\left(\frac{y}{z}\right)^{-(i+1)}-
\left(\frac{y}{z}\right)^i\right]\right)\cdot
\left(\sum_{j\in\BZ}B_{\beta+j}
B_{\beta-j}\left(\frac{y}{z}\right)^j+\sum_{i\neq
j\in\BZ}a_{ij}\frac{y^i}{z^j}\right)\\
&=&\frac{1}{yz}\left(\sum_{j\geq
0}b_j(B_{\beta+j}B_{\beta-j}-B_{\beta-j-1}B_{\beta+j+1})\right)+
\sum_{i{\rm ~or~}j\neq 0}c_{ij}y^{i-1}z^{j-1}
 \eean
and therefore, upon taking the double residue,
$$ \oint\limits_{\iy}\!\!\oint\limits_{\iy}
\frac{\rho_b^{(e)}(y/z)X(t;y)X(t;z)}{z^2(yz)^{\beta}}
 \frac{dy~dz}{(2\pi i)^2}= \sum_{j\geq
0}b_j(B_{\beta+j}B_{\beta-j}-B_{\beta-j-1}B_{\beta+j+1}).
$$
 \underline{For $N$ odd},
$$ \frac{X(t;y)X(t;z)}{(yz)^{\beta}
(y/z)^{1/2}}=\sum_{j\in\BZ}B_{\beta+\frac{1}{2}+j}B_{\beta-\frac{1}{2}-j}\left(
\frac{y}{z}\right)^j+\sum_{i\neq
j\in\BZ}a_{ij}\frac{y^i}{z^j} $$ and so

\medbreak\noindent
$\displaystyle{\rho_b^{(o)}\left(\frac{y}{z}\right)
\frac{X(t;y)X(t;z)}{z^2(yz)^{\beta}}}$ $$
=\frac{1}{yz}\sum_{j\geq
0}b_j\left(B_{\beta+j+\frac{1}{2}}B_{\beta-j-\frac{1}{2}}-B_{\beta-j-\frac{3}{2}
}B_{\beta+j+ \frac{3}{2}}\right)+\sum_{i{\rm
~or~}j\neq 0}c_{ij}y^{i-1}z^{j-1} $$ and therefore

$$ \oint\limits_{\iy}\!\!\oint\limits_{\iy}
\frac{\rho_b^{(o)}(y/z)X(t;y)X(t;z)}{z^2(yz)^{\beta}}
 \frac{dydz}{(2\pi i)^2} = \sum_{j\geq
0}b_j\left(B_{\beta+j+\frac{1}{2}}B_{\beta-j-\frac{1}{2}}-B_{\beta-j-\frac{3}{2}
}B_{\beta+j+ \frac{3}{2}}\right), $$
ending the proof of Proposition 5.2. \qed

\bigbreak

{\em Defining the set
 \be
 \BS_N^{(\ell)}:= \left\{
 \begin{array}{l}
  \displaystyle{\sg_1>\sg_2>\ldots >\sg_{\ell/2}, ~~\sg_i\in \BZ
  }\\ \\
 ~~~~~~~~
  \displaystyle{ \frac{\ell}{2}  \leq \sg_i+i \leq
   \left[\frac{N}{2}\right]}
   \end{array}
   \right\},
   \ee
 the map
\be
\sigma: ~\BY^{(\ell)}_{\frac{\ell(N-\ell)}{2}}
\longrightarrow
 \BS_N^{(\ell)}: ~\lb \longmapsto
  \sigma(\lb)=\left(
 \lb_i-i+\ell-\left[\frac{N+1}{2}\right] \right)
  _{1\leq i\leq n/2}.
 \label{sigma}\ee
is a bijection.

 }


Indeed, $\lb_1\geq \lb_2\geq \ldots$ implies at once
the strict inequalities $\sg_1> \sg_2 > \ldots$
 and also implies, together with the fact that for $\lb
 \in \BY^{(\ell)}_{\frac{\ell(N-\ell)}{2}}$ and $1\leq i\leq
 \ell/2$,
 $$
 2\lb_i\geq \lb_i+\lb_{\ell+1-i}=N-\ell~~\mbox{and, clearly} ~~
  \lb_i\leq N-\ell.
 $$
 Conversely, every $\sigma\in \BS_N^{(\ell)}$
 comes from a $\lb \in
 \BY^{(\ell)}_{\frac{\ell(N-\ell)}{2}}$.

\begin{lemma} For a given partition
$$ \lb=(\lb_1\geq\lb_2\geq
...\geq\lb_{\ell-2})\in\BY^{(\ell-2)}_{\frac{(\ell-2)(N-\ell+2)}{2}},
$$ and $j\geq 0$, the following holds
\be
B_{\beta+j}B_{\beta-j}\gs_{\lb}=-B_{\beta-j-1}B_{\beta+j+1}\gs_{\lb}=\left\{
\begin{array}{ll}
0,&\mbox{if $\beta+j=$ some $\lb_{\nu}-\nu-1$}\\
&\mbox{for $1\leq\nu\leq\ell/2-1$,}\\ &\mbox{or if $j
\geq N/2$}
\\
 \\
\gs_{\lb'},&\mbox{if $\beta+j\neq$ every
$\lb_{\nu}-\nu-1$}\\ &\mbox{for
$1\leq\nu\leq\ell/2-1$,}
\end{array}
\right. \label{A} \ee where
 \bea
 \lb'=\Bigl(\lb_1-2&\geq &
...\geq\lb_{\nu}-2\geq\beta+j+\nu\geq\lb_{\nu+1}-1\geq
...\geq\lb_{\frac{\ell}{2}-1}-1
 \nonumber\\
&\geq&\lb_{\frac{\ell}{2}}-1\geq ...\geq
\lb_{\ell-2-\nu}-1\geq(N-\ell)-(\beta+j+\nu)
\nonumber\\ &\geq&
 \lb_{\ell-1-\nu}\geq ...\geq\lb_{\ell-2}\Bigr)
\in\BY^{(\ell)}_{\frac{\ell(N-\ell)}{2}}.\label{B}
\eea
Moreover for $j$'s such that $\beta+j\neq$ every
$\lb_{\nu}-\nu-1$, the maps $B_{\beta+j}B_{\beta-j}$
induce maps 
\be
B_{\beta+j}B_{\beta-j}~~:~~
 \BY^{(\ell-2)}_{\frac{(\ell-2)(N-\ell+2)}{2}}
 \longrightarrow
\BY^{(\ell)}_{\frac{\ell(N-\ell)}{2}}~:~
\lb\longmapsto \lb' \label{Y-map}\ee
having, as a whole, a ``\underline{surjectivity
property}", meaning that to each $\lb' \in
\BY^{\ell}_{{(\ell-2)(N-\ell)}/{2}}$, there are
$\ell/2$ choices of $j\geq 0$ and
 $\lb \in
\BY^{(\ell-2)}_{{(\ell-2)(N-\ell+2)}/{2}}$ mapping to
$\lb'$,
by means of the map $B_{\beta+j}B_{\beta-j}$, as in (5.0.12).

At the level of the $\BS$-spaces, the maps
$B_{\beta+j}B_{\beta-j}$ induce maps
 \be
 \BS_N^{(\ell-2)} \longrightarrow \BS_N^{(\ell)}
:
 \sg=(\sg_1,\ldots, \sg_{\frac{\ell-2}{2}}) \longmapsto
  \sg'=(\sg_1,\ldots,\sg_{\nu},j,\sg_{\nu+1},\ldots,
   \sg_{\frac{\ell-2}{2}}),
   \label{S-map}\ee
 having the same ``{surjectivity
property}" as above.

\bigbreak

\underline{For $N$ odd}, all formulae above remain the
same , except for the substitution $j\mapsto
j+\frac{1}{2}$ in (\ref{A}) and (\ref{B}).

\end{lemma}

\proof Extending a classic identity (see MacDonald
\cite{MacDonald})
 to arbitrary sequences $(\lb_1,...,\lb_n)$, we have
$$
B_{\lb_1},...,B_{\lb_n}(1)=(\lb_1,...,\lb_n):=\det\left(\gs_{\lb_i+j-i}(t)\right
)_{1\leq i,j\leq n} $$ and, in particular, for a
partition $(\lb_1\geq\lb_2\geq ...\geq\lb_{\ell})$, we
have, for an arbitrary choice of $j\geq 0$,
\bea
B_{\beta+j}B_{\beta-j}\gs_{(\lb_1,...,\lb_{\ell-2})}&=&\gs_{(\beta+j,\beta-j,
\lb_1,...,\lb_{\ell-2})}\no\\
&=&\det\left(\begin{array}{ccccc}
\gs_{\beta+j}&\gs_{\beta+j+1}&\gs_{\beta+j+2}&...&\gs_{\beta+j+\ell-1}\\
\gs_{\beta-j-1}&\gs_{\beta-j}&\gs_{\beta-j+1}&...&\gs_{\beta-j+\ell-2}\\
\gs_{\lb_1-2}&\gs_{\lb_1-1}&\gs_{\lb_1}&...&\gs_{\lb_1+\ell-3}\\
\vdots& & &\ddots&\vdots\\
\gs_{\lb_{\ell-2}-\ell+1}&...&...&...&\gs_{\lb_{\ell-2}}
\end{array}\right).
\no\\ \label{Schurmatrix}
  \eea
  Using the value (\ref{samebeta}) of $\beta$, it is
immediately clear, from the matrix (\ref{Schurmatrix}), that for
$j\geq {N/2}$, the second row of the matrix (\ref{Schurmatrix})
vanishes and therefore the determinant. Therefore we assume $0\leq
j\leq\frac{N}{2}-1$. We give the proof for even $N$; for odd $N$,
it is identical with $j \mapsto  j+{1}/{2}$.

The first column of the matrix above involves the
indices
\be
\frac{N}{2}-\ell+1+j,\frac{N}{2}-\ell-j,\lb_1-2,\lb_2-3,...,\lb_{\frac{\ell}{2}}
-\frac{\ell}{2}-1,...,\lb_{\ell-2}-\ell+1.
 \label{indices}\ee
 Consider now an arbitrary integer $j\geq 0$ and an arbitrary
partition
$$
\lb\in\BY^{(\ell-2)}_{\frac{(\ell-2)(N-\ell+2)}{2}};
$$
 it has the property that
$$ \lb_i+\lb_{\ell-1-i}=N-\ell+2\mbox{~for~}1\leq
i\leq
 \frac{\ell-2}{2}.
  $$
  Hence, for $i=\frac{\ell-2}{2}$
$$
2\lb_{\frac{\ell}{2}}\leq\lb_{\frac{\ell}{2}-1}+\lb_{\frac{\ell}{2}}=N-\ell+2,
$$ and so

$$ \lb_{\ell/2}\leq\frac{N-\ell+2}{2}; $$ thus, for
the arbitrary $j\geq 0$ chosen above

$$ \lb_{\ell/2}-\ell/2 -1\leq\frac{N}{2}-\ell
<\frac{N}{2}-\ell+j+1. $$ The partition
$\lb_1\geq\lb_2\geq ...$ implies the strict
inequalities
$$ \lb_1-1-1>\lb_2-2-1>\lb_3-3-1>\ldots >\lb_{\nu
+1}-(\nu+1)-1>\ldots
>\lb_{\ell/2}-\ell/2-1
$$ and therefore, there exist
$0\leq\nu\leq\frac{\ell}{2}-1$ such that

$$ \lb_{\nu}-\nu-1\geq\frac{N}{2}-\ell
+j+1\geq\lb_{\nu+1}-\nu-2. $$ These inequalities
together with the fact that

$$ \lb_{\nu}+\lb_{\ell -1-\nu}=N-\ell
+2,\quad\lb_{\nu+1}+\lb_{\ell -2-\nu}=N-\ell+2 $$ also
imply
 $$ \lb_{\ell -2-\nu}-(\ell
-1-\nu)\geq\frac{N}{2}-\ell -j\geq\lb_{\ell
-1-\nu}-(\ell-\nu).
 $$
 Therefore,
 the indices (\ref{indices}) of the first column of
 the matrix (\ref{Schurmatrix}) are now
  rearranged by order, as follows:
\bea &
&\lb_1-2>\lb_2-3>\ldots>\lb_{\nu}-\nu-1\geq\frac{N}{2}
 -\ell+1+j\geq\lb_{\nu+1}-\nu -2
 \nonumber\\
  &&
>\ldots>\lb_{\frac{\ell}{2}-1}-\frac{\ell}{2}>
 \lb_{\frac{\ell}{2}}-\frac{\ell}{2}-1
  >\ldots>\lb_{\ell-2-\nu}-(\ell-1-\nu)
   \nonumber\\
&&\geq \frac{N}{2}-\ell-j\geq
\lb_{\ell-1-\nu}-(\ell-\nu)>\ldots>\lb_{\ell-2}-\ell+1.
\label{5.0.4}
 \eea

Notice that the determinant (\ref{Schurmatrix}) vanishes if any of
the equalities hold in (\ref{5.0.4}) above. Therefore we may
assume strict inequalities. Upon rearranging the rows of the
matrix (\ref{Schurmatrix}) according to the order in
(\ref{5.0.4}}), we now list the corresponding partitions by
looking at the indices on the diagonal. This amounts to adding
$i-1$ to the $i^{\rm th}$ entry of (\ref{5.0.4}), thus leading to
$$ {\scriptsize \begin{array}{lcclccclc}
\lb'=&\Big(\lb_{1}-2\geq&\lb_{2}-2&\geq
\ldots\geq&\lb_{\nu}-2\geq&\frac{N}{2}-\ell+1+j+\nu\geq&
\lb_{\nu+1}-1&\geq\ldots\geq&\lb_{\frac{\ell}{2}-1}-1\\
 &\stackrel{\uparrow}{1}&\stackrel{\uparrow}{2}&
 &\stackrel{\uparrow}{\nu}&\stackrel{\uparrow}{\nu
 +1}&\stackrel{\uparrow}{\nu +2}& &\stackrel{\uparrow}{\ell/2}
\end{array}
} $$ $$ {\scriptsize \begin{array}{lclccclc}
\geq&\lb_{\frac{\ell}{2}}-1&\geq\ldots\geq&\lb_{\ell-2-\nu}-1\geq&\frac{N}{2}-j-
\nu-1\geq&
\lb_{\ell-1-\nu}&\geq\ldots\geq&\lb_{\ell-2}\Bigr)\\
 &\stackrel{\uparrow}{\ell/2+1}& &\stackrel{\uparrow}{\ell
-2-\nu+1}&\stackrel{\uparrow}{\ell-2-\nu+2}&
\stackrel{\uparrow}{\ell-2-\nu +3}&
&\stackrel{\uparrow}{\ell}
\end{array}
} $$
 \be
  \label{5.0.5}
  \ee
The rearrangement does not change the sign of the determinant
(\ref{Schurmatrix}). Knowing that
$\displaystyle{\lb\in\BY^{(\ell-2)}_{\frac{(\ell-2)(N-\ell+2)}{2}}}$,
we now prove that the new partition $\lb'$ (obtained in
(\ref{5.0.5})),

$$ \lb'\in\BY_{\frac{\ell(N-\ell)}{2}}^{(\ell)}; $$
i.e., we prove
\bean {\rm (i)} \hspace{1cm}
\sum_i^{\ell}\lb'_i&=&\sum_1^{\ell-2}\lb_i-2\nu-
(\ell-2-2\nu)+\left(\frac{N}{2}-\ell+1+j+\nu \right)
\\&&\hspace{6cm}+\left(\frac{N}{2}-j-\nu-1\right)
 \\&=&\frac{\ell(N-\ell)}{2},
  \eean
 and
  $$ \lb'_i+\lb'_{\ell+1-i}=N-\ell\quad\mbox{for
all $1\leq
i\leq\displaystyle{\frac{\ell}{2}}$;}\leqno{{\rm
(ii)}}
 $$
 e.g.,
 \bean
\lb'_i+\lb'_{\ell+1-i}&=&\lb_i-2+\lb_{\ell-1-i}=N-\ell\quad\mbox{for
$1\leq i\leq\nu$}\\
 & & \\
\lb'_{\nu+1}+\lb'_{\ell-\nu}&=&\left(\frac{N}{2}-\ell+1+j+\nu
\right)+\left(\frac{N}{2}-j-\nu-1\right)=N-\ell\\&&\\
\lb'_i+\lb'_{\ell+1-i}&=&\lb_{i-1}-1+\lb_{\ell-i}-1=
 N-\ell\quad\mbox{for
$\nu+2\leq i\leq\ell/2$}
 . \eean
  So far, we have shown that to
an arbitrary integer $j\geq 0$ and a partition
$$ \lb=(\lb_1\geq\lb_2\geq
...\geq\lb_{\ell-2})\in\BY_{\frac{(\ell-2)(N-\ell+2)}{2}}^{(\ell-2)},
$$
 such that the inequalities in (\ref{5.0.5}) are
strict, corresponds a new partition
$$ \lb'=(\lb'_1\geq ...\geq\lb'_{\ell})\in
\BY_{\frac{\ell(N-\ell)}{2}}^{(\ell)},
 $$
  with $\lb'$
totally determined by (\ref{5.0.5}). Then $\ell/2$ different
choices of $\lb\in\BY_{{(\ell-2)(N-\ell+2)}/{2}}^{(\ell-2)}$ and
$j\geq 0$ will lead to the same sequence of numbers (\ref{5.0.5}),
as appears from the next argument.

In view of the $\sg$-map in (\ref{sigma}), it is obvious to see
that the $\nu+1$-st number in $\lb'$ of (\ref{5.0.5}) gets mapped
by $\sg$ into $j$, namely
 $$
  \frac{N}{2}-\ell+1+j+\nu \longmapsto
  j,
  $$
  and, in general, (5.0.15) holds.
  The ``surjectivity property" is
  straightforward in this description, since given a
  sequence $\sg' \in \BS_N^{(\ell)} $, you may choose $j$ to be any
  of the $\ell/2$ numbers appearing in $\sigma'$; then
  $\sigma$ is the sequence formed by the remaining numbers in order.
This establishes Lemma 5.3.%
\qed

\begin{proposition
 } Given positive integers $N$ and $\ell$ with even $\ell$
and the operator $$ Y_{\beta}=\left\{
\begin{array}{l}
\displaystyle{\sum_{j\geq 0}b_j\left(B_{\beta
+j}B_{\beta -j}-B_{\beta -j-1}B _{\beta +j+1}\right)},
~~N\mbox{~even,}\\
 \\
\displaystyle{\sum_{j\geq 0}b_j\left(B_{\beta
+j+\frac{1}{2}}B_{\beta -j-\frac{1}{2}}-B_{\beta
-j-\frac{3}{2}}B _{\beta +j+\frac{3}{2}}\right)},
~~N\mbox{~odd,}
\end{array}
\right. $$ we have $$
Y_{\frac{N}{2}-\ell+1}\tau_{\ell-2}=\ell\tau_{\ell}.
$$
\end{proposition}

\proof The indices of the $b_i$ in the $\ell$th
tau-function can now be expressed in terms of the
$\sg$-map, as follows
 $$
   \tau_{\ell}(t)
 =
  \sum_{\lb\in\BY^{(\ell)}_{\frac{\ell
(N-\ell)}{2}}}\left(\prod_1^{\ell/2}b_{\lb_
i-i+\ell-\left[\frac{N+1}{2}\right]} \right){\bf
s}_{\lb}(t)
 =
 \sum_{\lb\in\BY^{(\ell)}_{\frac{\ell
(N-\ell)}{2}}}\left(\prod_1^{\ell/2}b_{\sg_ i(\lb)}
\right){\bf s}_{\lb}(t)
 . $$
We give the proof for even $N$.
%
From (\ref{S-map}), it follows at once that
  \be
   b_j \prod_1^{(\ell-2)/2}b_{\sg_ i(\lb)}=
  \prod_1^{\ell/2}b_{\sg_ i(\lb')}.
  \label{o}\ee
%
%
Setting $\displaystyle{Y_{\beta}=\sum_{i\geq 0}b_i\Gamma_i}$, one
computes, using Lemma 5.3, (\ref{o}) and in $\stackrel{*}{=}$ the
$\ell/2$-to-$1$ ``surjectivity" of the maps (\ref{Y-map}) or
(\ref{S-map}):

\bean
Y_{\frac{N}{2}-\ell+1}\tau_{\ell-2}(t)
 &=&\sum_{\lb\in\BY^{(\ell-2)}_
{\frac{(\ell-2)(N-\ell-2)}{2}}
}\left(\prod_1^{\frac{\ell-2}{2}}b_{\sg_i(\lb)
 }\right)Y_{\beta}(
s_{\lb}(t))\\
&=&
 \sum_{\lb\in
 \BY^{(\ell-2)}_{\frac{(\ell-2)(N-\ell-2)}{2}}}\sum_{j\geq
0} \left(\prod_1^{\frac{\ell-2}{2}} b_{\sg_
i(\lb)}\right)b_j\Gamma_j (s_{\lb}(t))\\
&\stackrel{*}{=}&\frac{\ell}{2}
 \sum_{\lb'\in\BY^{\ell}_{\frac{\ell(N-\ell)}{2}}}
\prod^{\ell/2}_1 b_{\sg_ i(\lb')}2s_{\lb'}(t)\\
&=&\ell\tau_{\ell}(t),
 \eean
  ending the proof of Proposition 5.4.
   \qed

\medbreak

\noindent {\it Proof of Theorem 5.1:} Formula (\ref{statementA})
follows at once from Propositions 5.2 and 5.4. To prove
(\ref{statementB}), first notice that, upon setting
$I_{\ell}:=\displaystyle{\left(\frac{\ell}{2}\right)}!\tau_{\ell}$,
$$
(\BX(t;y)\BX(t;z)I)_{\ell}=y^{\ell-1}z^{\ell-2}X(t;y)X(t;z)I_{\ell-2}.
$$ Then

\bean (\BY(t) I)_{\ell}&=&\left(\oiint
 \BX(t;y)\BX(t;z)\frac{\rho_b(y/z)dy~
   dz}{2z(yz)^{N/2}}I\right)_{\ell}\\
 & &\\
 &=&\oiint
   \frac{dy~dz~\rho_b(y/z)}{2z^2(yz)^{N/2-\ell+1}}
 X(t;y)X(t;z)I_{\ell-2}\\ & &\\
 &=&\frac{1}{2}Y_{\frac{N}{2}-\ell+1}I_{\ell-2}
 ,~~\mbox{by definition (5.0.5) of $Y_{\beta}$,}\\ & &\\
&=&\frac{1}{2}Y_{\frac{N}{2}-\ell+1}
\left(\frac{\ell-2}{2}\right)!~\tau_{\ell-2}\\ & & \\
&=&\left(\frac{\ell}{2}\right)!~\tau_{\ell},~ \mbox{
using (\ref{statementA})}\\ & &
\\ &=&I_{\ell}, \eean
 ending the proof of Theorem 5.1. \qed

 \bigbreak


\noindent{\bf Example}:
 {\sl For $b_i=2i+1$ and even $N$, the function
$\rho_b(x)$, defined in (\ref{rho1}), equals\footnote{$\rho_b(x)$
is actually a distribution! }
\be
  \rho_b(x)=\sum_{i\geq
0}b_i(x^{-i-1}-x^i)=-\frac{1+x}{(1-x)^2}+x^{-1}\frac{1+x^{-1}}{(1-x^{-1})^2}
.\label{rho for simple b}
 \ee
 The corresponding vertex operator (\ref{Y-operator})
 takes on a particularly simple form:
 \be
   Y_{\frac{N}{2}-\ell +1}
  =
  2B^{(2)}_{N-2\ell+2}  =
 2\int\limits_{\BR}du~\delta^{(N-2)}(u)u^{2\ell-4}
  X^{(2)}(u),\label{Y for simple b}
 \ee
  where $\delta^{(N-2)}$ is the $(N-2)$nd derivative
  of the customary $\delta$-function and
  where the $B^{(2)}_i $ are the differential operators
  (\ref{B-operators}) in the $t_i$,
  $$
  B^{(2)}_i:=\sum_{j\geq 0} {\bf s}_{i+j}(2 t) {\bf s}_j(-2
  \tilde \pl_{t}),
  $$ given by
  the coefficients of the expansion in
  powers of $z$ of the vertex operator
$$ X^{(2)}(z):=e^{2\sum_1^{\iy}t_iz^i}e^{-2\sum_1
^{\iy}\frac{z^{-i}}{i}\frac{\pl}{\pl t_i}}
 =
 \sum_{i \in \BZ} B^{(2)}_i z^i. $$
}
%

\proof  Formula (\ref{rho for simple b}) follows immediately from
the series
 $$
 \frac{1+x}{(1-x)^2}=1+3x+5x^2+7x^3+\ldots
~~. $$
Setting, for convenience,
  $$
X(t;y,z):=e^{\sum_1^{\iy}t_i(y^i+z^i)}
 e^{-\sum_1^{\iy}
\left(\frac{y^{-i}+z^{-i}}{i}\right) \frac{\pl}{\pl
t_i}}
 $$
  and using
$X(t;y)X(t;z)=\displaystyle{\left(1-\frac{z}{y}\right)}
 X(t;y,z)$ and $X(t;z,z)=X^{(2)}(t;z)$,
one computes ($\beta=\frac{N}{2}-\ell +1$)
\bean Y_{\beta}&=&\oiint
\frac{\rho^e(y/z)}{(yz)^{\beta}z^2}X(t;y)X(t;z) dy~dz\\
&=&\oiint
 \left(
 \frac{z(1+z/y)}{y(1-z/y)^2}-
\frac{(1+y/z)}{(1-y/z)^2}\right)\frac{(1-z/y)}{z^2(zy)^{\beta}}
 X(t;y,z)dy~dz\\
&=&\oiint
\frac{z}{y}\frac{(1+z/y)}{(1-z/y)^2}\frac{(1-z/y)}{z
^2(zy)^{\beta}} X(t;y,z)dy~dz\\
&=&\oint\limits_{\iy}\left(\oint\limits_{\iy}
 \frac{(1+z/y)}{(y-z)z(zy)^{\beta}}X(t;y,z)
  \frac{dy}{2\pi i}\right)\frac{dz}{2\pi i}\\
&=&2\oint\limits_{\iy}\frac{X(t;z,z)}{z^{2\beta+1}}\frac{dz}{2\pi
i}
\\
&=&2\oint\limits_{\iy}\frac{X^{(2)}(t;z)}{z^{2\beta+1}}
 \frac{dz}{2\pi i}
  \\
  &=&2B^{(2)}_{2\beta}=
2B^{(2)}_{N-2\ell+2}=2\int\limits_{\BR}du~
 \delta^{(N-2)}(u)u^{2\ell-4}X^{(2)}(t;u),
\eean
 establishing (\ref{Y for simple b}).\qed


\section{Duality}

\begin{proposition}For odd $N$ and odd $\ell$, the
following holds:

\bea
 \tilde\tau_{\ell}(t)&:=&
z^{-1}\det {}^{1/2}\left( E_{\ell,N}(t)~\Bigl( m_N(0)
 +z^2 \vr_{\frac{N+1}{2},
 \frac{N+1}{2}}\Bigr)~
    E^{\top}_{\ell,N}(t)\right)\nonumber\\
    &=&
      \sum_{\lb\in\BY^{(\ell)}_{\frac{\ell
(N-\ell)}{2}}}\left(\prod_1^{[\ell/2]}b_{\lb_
i-i+\ell-\left[\frac{N+1}{2}\right]} \right){\bf
s}_{\lb_1 \geq \ldots \geq \lb_{\ell}}(t).
 \eea
Then these functions
 \be
  \tilde \tau_{\ell}(t)=
(-1)^{\ell(N-\ell)/2} ~ \left(\prod_0^{\frac{N-3}{2}}b_i\right)
\left( \left.\tau_{N-\ell}(-t)\right|_{b_i\rightarrow
 b_i^{-1}}\right),~~~\mbox{for $\ell$ odd}.
\ee
 are the $\tau$-functions $\tau_k(t)$ (in reverse order
  and modulo a multiplicative factor) of
 the Pfaff lattice for odd $N$ and even $k$, with
 $t\mapsto -t$, and with initial condition
\be
 \left(
\begin{array}{ccccccc}
 &&O& & & &b^{-1}_{\frac{N-3}{2}}\\
 && & & &\diagup& \\
 && & & b^{-1}_0& & \\
 && &0& & & \\
&& -b^{-1}_0&& & & \\
 &\diagup&& & & & \\
-b^{-1}_{\frac{N-3}{2}}&& & & &O&
\end{array}
\right). \ee

 \end{proposition}

 \proof
Defining $k_i$ and $k^{\top}_i$ by \be
\lb_i=k_i-\ell+i,\qquad\lb_i^{\top}=k^{\top}_i-(N-\ell)+i,
\label{4}\ee it is easy to see the one-to-one correspondence
between
$$ \BY^{(\ell)}_{\frac{\ell(N-\ell)}{2}} \df
\left\{\begin{array}{l} N-1\geq
k_1>k_2>...>k_{\ell}\geq 0\\
\\
\mbox{with $k_i+k_{\ell+1-i}=N-1$ for $1\leq
i\leq\frac{\ell+1}{2}$}
\end{array}\right\}
$$ and also between

\be \BY^{(N-\ell)}_{\frac{\ell(N-\ell)}{2}}
\df\left\{\begin{array}{l} N-1\geq
k_1^{\top}>k_2^{\top}>...>k_{N-\ell}^{\top}\geq
0\\
 \\
\mbox{with $k_i^{\top}+k^{\top}_{N-\ell+1-i}=N-1$ for $1\leq
i\leq\frac{N-\ell}{2}$}
\end{array}\right\}.
\ee

\begin{lemma}

{\rm (1)}  The following correspondence holds
 \be
  \lb \in
\BY^{(\ell)}_{\frac{\ell(N-\ell)}{2}}
 \df
   \lb^{\top}\in \BY^{(N-\ell)}_{\frac{\ell(N-\ell)}{2}},
   \ee

{\rm (2)} For $\lb$ and $\lb^{\top}$, we have the following
disjoint union
 \be
 \{k_1>...>k_{\ell}\}\cup\{k_1^{\top}>...>k^{\top}_{N-\ell}\}
 =\{0,1,...,N-1\}.
 \ee

\end{lemma}

\proof Considering

$$ (\lb_1\geq\lb_2\geq ...\geq\lb_{\ell})\in
\BY^{\ell}_{\frac{\ell(N-\ell)}{2}}, $$ we have \bea &
&\lb^{\top}_1=...=\lb^{\top}_{\lb_{\ell}}=\ell\nonumber\\
 & & \nonumber\\
&
&\lb^{\top}_{\lb_{\ell}+1}=...=\lb^{\top}_{\lb_{\ell-1}}=\ell-1\nonumber\\
& & \nonumber\\ &
&\lb^{\top}_{\lb_{\ell-1}+1}=...=
\lb^{\top}_{\lb_{\ell-2}}=\ell-2   \label{8}\\
& & \nonumber\\ & &\quad \vdots\nonumber
 \eea
  and so, since
\bean k_i=N-1-k_{\ell
+1-i}&=&N-1-(\lb_{\ell+1-i}+\ell-(\ell+1-i))\\
&=&N-(\lb_{\ell+1-i}+i)
 \eean
 we have, on the one hand
 \be
  k_1=N-\lb_{\ell}-1>k_2=N-\lb_{\ell-1}-2>k_3=N-\lb_{\ell-2}-3,
 \label{9}
  \ee
  and on the other hand, using (\ref{4}) and (\ref{8}),
{\footnotesize \bea
 &&\hspace{-1cm}
 k^{\top}_1=N-1>k_2^{\top}=N-2>\ldots>k_{\al}^{\top}=N-\al>\ldots
  >k^{\top}_{\lb_{\ell}}=N-\lb_{\ell}>\nonumber\\
& &\hspace{-1cm} k^{\top}_{\lb_{\ell}+1}=N-\lb_{\ell}-2>
\ldots>k_{\beta}^{\top}=N-\beta-1  >\ldots
>k^{\top}_{\lb_{\ell-1}}=N-\lb_{\ell-
1}-1>\nonumber\\
 & &\hspace{-1cm}
 k^{\top}_{\lb_{\ell-1}+1}=
N-\lb_{\ell-1}-3>\ldots>k_{\gamma}^{\top}=N-\gamma-2
>\ldots>k^{\top}_{\lb_{\ell-2}}=
 N-\lb_{\ell-2}-2>\ldots\nonumber\\
& & \label{10}\eea}
\noindent So the gaps in (\ref{10}) coincide with the sequence
(\ref{9}). This ends the proof of Lemma 6.2.\qed

\bigbreak

\noindent{\it Proof of Proposition 6.1:} One checks, using
Proposition 3.2, 
%
\bean
 \tilde\tau_{\ell}(t)
 &=&\sum_{\lb\in\BY^{(\ell)}_{\frac{\ell(N-\ell)}{2}}}
\prod_1^{\frac{\ell-1}{2}}
 b_{\lb_i-i+\ell-\frac{N+1}{2}}~{\bf s}_{\lb}(t)\\
 \\
&=&(-1)^{|\lb|}\sum_{\lb\in\BY^{(\ell)}_{\frac{\ell(N-\ell)}{2}}}
\prod_1^{\frac{\ell-1}{2}}b_{k_i-\frac{N+1}{2}}
 ~{\bf s}_{\lb^{\top}}(-t)\\
 \\
&=&(-1)^{|\lb|}\prod_0^{\frac{N-3}{2}}b_i\sum_{\lb\in\BY^{(\ell)}_{\frac{\ell(N-
\ell)}{2}}} \frac{1}{\displaystyle{\prod_1^{\frac{N-\ell}{2}}
}b_{k_i^{\top}-\frac{N+1}{2}}}~{\bf
s}_{\lb^{\top}}(-t)~,~~\mbox{using Lemma 6.2,}\\
 \\
&=&(-1)^{|\lb|}\prod_0^{\frac{N-3}{2}}b_i\sum_{\lb^{\top}\in\BY^{(N-\ell)}
_{\frac{\ell(N-\ell)}{2}}}\prod_1^{\frac{N-\ell}{2}}b^{-1}_{\lb^{\top}_i
-i+\ell-\frac{N+1}{2}}~{\bf s}_{\lb^{\top}}(-t)\\
 \\
&=&
 (-1)^{\frac{\ell(N-\ell)}{2}}\prod_0^{\frac{N-3}{2}}b_i\left(\tau_{N-\ell}(-t
) \Big|_{b_i\rg b_i^{-1}}\right),
 \eean
 which is, using Theorem
1.1, the $\tau$-function (modulo a constant) for the case where
$N$ is odd and $N-\ell$ even, concluding the proof of the
Proposition.  \qed


\section{Examples}

\subsubsection*{Example 1: Rectangular Jack
polynomials}

\begin{proposition} When
\bea
 b_i&=&2i+1~~\mbox{for $N$ even}\nonumber\\
    &=& 2i+2 ~~ \mbox{for $N$ odd},
    \eea
 then the $\tau_{2n}(t)$'s are Jack polynomials for
rectangular partitions, with $n\leq [N/2]$,
 \bea
  \tau_{2n}(t)&=&
  pf~m_{2n}(t)\nonumber\\
&=&
 \sum_{\lb\in\BY^{(2n)}_{{n(N-2n)}}}\prod_1^{n}
  (k_i-k_{2n+1-i}
){\bf s}_{\lb}(t), \mbox{~where~}k_i=\lb_i-i+2n
%
 \nonumber\\ & & \nonumber\\
 &=&J^{(1/2)}_{\lb}(x)\Bigg|_{t_i=\frac{1}{i}
\displaystyle{\sum_k}x^i_k}\quad\mbox{for the
partition~}\lb
=(N-2n)^n
\nonumber\\
&=&\frac{1}{n!}\int_{\BR^n}\Delta(z)^4\prod_{k=1}^n
 e^{2\displaystyle{\sum^{\iy}_
1}t_i z^i_k}\delta^{(N-2)}{(z_k)}dz_k.
 \eea
Then $$
   m_{\ell}(t)=E_{\ell,N}(t) m_N(0)
    E^{\top}_{\ell,N}(t),
    $$
with
  \bean m_N(0)&=&{\tiny \left(
\begin{array}{cccccccc}
& &O& & & &&N-1\\
  & & & & & &N-3& \\
 & & & & &\diagup& & \\
 & & & & 1& & & \\
 & & &-1& & & \\
  &&\diagup& & & & \\
 & -N+3& & & & &  \\
 -N+1& & & & & & O
\end{array}
\right) }  ~~\mbox{,  for $N$ even,} \nonumber \\ &&
 \nonumber\\
 m_N(0)&=&
  {\tiny \left(
\begin{array}{ccccccccc}
& &O& & && &&N-1\\
  & & & && & &N-3& \\
 & & & & &&\diagup& & \\
 & & & && 2& & & \\
 & & & &0& & & \\
 & & &-2& & & & \\
 & &\diagup&& & & & \\
 & -N+3&& & & & &  \\
 -N+1& & & & & & & O
\end{array}
\right)}
  ~~\mbox{,  for $N$ odd}
   \eean  
where (setting
$\tilde{\bf s}_n(t)={\bf s}_n(2t)$)
  \bea
\lefteqn{m_N(t)}\nonumber\\
 &=&
  \left((j-i)\tilde{\bf s}_{N-i-j-1}
 \right)_{0\leq i,j\leq N-1}
  \nonumber\\ && \nonumber\\
&=&
 \footnotesize{
\left(\begin{array}{cccccccccc} 0&\tilde{\bf s}_{N-2}&
& 2\tilde{\bf s}_{N-3}& &\ldots& (N-2)\tilde{\bf s}_1&
&N-1\\
\\
-\tilde{\bf s}_{N-2}&0& &\tilde{\bf s}_{N-4}&
&\ldots&N-3\\
\\
-2\tilde{\bf s}_{N-3}&-\tilde{\bf s}_{N-4}&
&\quad\quad 0& &\diagup\\
\\
 & & & &1\\
\vdots&\vdots& &\quad\quad -1\\
\\
& &\quad \diagup\\
\\
-(N-2)\tilde{\bf s}_1&-N+3\\
\\
 & & & & & & &O\\
-N+1
\end{array}\right)
 }
 \nonumber\\
&&\hspace{8.9cm} \mbox{for $N$ even}\nonumber\\ &&
 \nonumber\\
 &=&
 \footnotesize{\left(\begin{array}{cccccccccc} 0&\tilde{\bf
s}_{N-2}& & 2\tilde{\bf s}_{N-3}& &\ldots&
(N-2)\tilde{\bf s}_1& &N-1\\
\\
-\tilde{\bf s}_{N-2}&0& &\tilde{\bf s}_{N-4}&
&\ldots&N-3\\
\\
-2\tilde{\bf s}_{N-3}&-\tilde{\bf s}_{N-4}&
&\quad\quad 0& &\diagup\\
\\
 & & & &2\\
 & & &0 &  \\
\vdots&\vdots&-2 &   \\
\\
& & \diagup\quad\quad\quad\\
\\
-(N-2)\tilde{\bf s}_1&-N+3\\
\\
 & & & & & & &O\\
-N+1
\end{array}\right)
 }
\nonumber\\
&&\hspace{8.9cm} \mbox{for $N$ odd}\nonumber\\ \eea
\end{proposition}

\proof Setting $$
t_k=\frac{1}{k}\sum^{\ell}_{i=1}x_i^k $$ we have \bean
e^{\beta\displaystyle{\sum_1^{\iy}}t_kz^k}&=&e^{\beta
\displaystyle{\sum_{i=1}^{\ell}\sum_{k=1}^{\iy}\frac{1}{k}}(x_iz)^k}\\
&=&\prod^{\ell}_{i=1}\left(e^{\displaystyle{\sum_{k=1}^{\iy}\frac{1}{k}}(x_iz)^k
}\right)^{\beta}\\
&=&\prod^{\ell}_{i=1}(1-x_iz)^{-\beta} \eean

\medbreak

According to Awata et al. \cite{awata}, the Jack polynomials for
rectangular partitions $s^n$ have the following integral
representation: (for connections with random matrix theory, see
\cite{PvM})

\bean
cJ_{s^n}^{1/\beta}&=&\oint_{z_1=...=z_n=0}\vert\Delta(z)
 \vert^{2\beta}\prod^n_{j=1}z_j^{-(n-1)\beta -s}
\prod^{\ell}_{i=1}(1-x_iz_j)^{-\beta}
 \frac{dz_j}{2\pi iz_j}  \\
&=&\oint_{z_1=...=z_n=0}\vert\Delta(z)\vert^{2\beta}\prod^n_{j=1}
 z_j^{-(n-1)\beta -s}
e^{\beta\displaystyle{\sum_{k=1}^{\iy}}t_kz^k_j}\frac{dz_j}{2\pi
iz_j}\\
&=&c_n\int_{\BR^n}\vert\Delta_n(z)\vert^{2\beta}
\prod^n_{j=1}
 e^{\beta \displaystyle{\sum_{k=1}^{\iy}}t_kz^k_j}
 \delta^{s+(n-1)\beta}(z_j)dz_j.
\eean

\medbreak

\newpage

Setting $\beta=2,~s=N-2n$ and $2\leq 2n \leq N$ in the last
integral , we have, using the standard derivation of the
``symplectic" matrix integral, (see \cite{AvM6})

\bean
\lefteqn{\frac{1}{n!}\int_{\BR^n}\Delta^4_n(z)
 \prod_{k=1}^n e^{2\sum_{k=1 }^{\iy} t_kz^k_j}
  \delta^{N-2}(z_j)dz_j}\\
&=&
 pf\left( \int_{\BR}\{y^k,y^{\ell}\}e^{2
 \sum_{i=1}^{\iy}
 t_iy^i}\delta^{(N-2)}(y)dy\right)_{0\leq
k,\ell\leq 2n-1}\\
&=&pf\left((k-\ell)\int_{\BR}y^{k+\ell-1}e^{2
 \sum_{i=1}^{\iy}
 t_iy^i}\delta^{(N-2)}
(y)dy\right)_{0\leq
k,\ell\leq 2n-1}\\
&=&pf\left((k-\ell)\sum_{i=0}^{\iy}\tilde{\bf
s}_i(t)\int_{\BR}
 y^{i+k+\ell
-1}\delta^{(N-2)}(y)dy\right)\\
\\
&=&
 pf\Bigl((-1)^{N-2}(N-2) !(k-\ell)\tilde{\bf s}_{N-1
 -k-\ell}(t)\Bigr)_{0\leq k,\ell\leq 2n-1}\\
\\
&=&
c_{N,n}pf \Bigl((\ell-k)\tilde{\bf s}_{N-1-k-\ell}(t)\Bigr)_{0\leq
k,\ell\leq 2n-1}. \eean\vspace{-1.3cm} \be \ee

In order to find the initial condition $m_N(0)$, one sets $t=0$ in
the last matrix appearing in (7.0.3), to yield
 $$
 \Bigl((\ell-k)\tilde{\bf
s}_{N-1-k-\ell}(0)\Bigr)_{0\leq k,\ell\leq N-1}.
 $$
All entries of this matrix vanish, except the
antidiagonal, from which one reads off the $b_i$'s:

\noindent \underline{For $N$ even}, we have $b_i=2i+1$
and thus
 \bean
 b_{\lb_i-i+\ell-N/2}&=&
2\left(\lb_i-i+\ell-\frac{N}{2} \right)+1
\\
&=&\lb_i-\lb_{\ell+1-i}-2i+\ell+1~~
\mbox{~using~}\lb_i+\lb_{\ell+1-i}=N-\ell\\
 &=&  k_i-k_{\ell+1-i}~~\mbox{~using~} k_i=\lb_i-i+2n.
 \eean

\noindent \underline{For $N$ odd}, we have $b_i=2i+2$
and thus
  \bean
 b_{\lb_i-i+\ell-(N+1)/2}&=&
2\left(\lb_i-i+\ell-\frac{N+1}{2} \right)+2
\\
&=&\lb_i-\lb_{\ell+1-i}-2i+\ell+1~~
\mbox{~using~}\lb_i+\lb_{\ell+1-i}=N-\ell\\
 &=&  k_i-k_{\ell+1-i},
 \eean
ending the proof of Proposition 7.1.\qed

 \newpage
 \begin{example} For $n=4$ and $b_0=1,~b_1=3$, the solution to the system
(1.0.8) is given by \be
L=\frac{1}{(t_2+t_1^2)^2}\MAT{4} 0&1&0&0\\
\\
t_1&2(t_2-t_1^2)&-\sqrt{3}t_1&0\\
\\
\frac{2}{\sqrt{3}}(t_2-t^2_1)&-\frac{16}{\sqrt{3}}t_1t_2&-2(t_2-t_1^2)&1\\
\\
-\sqrt{3}t_1&-2\sqrt{3}(t_2-t_1^2)&3t_1&0 \mat . \ee
Indeed $$ m_4=\MAT{4} 0&-\tilde{\bf s}_2&-2\tilde{\bf
s}_1&-3\\ \tilde{\bf s}_2&0&-1&0\\ 2\tilde{\bf
s}_1&1&0&0\\ 3&0&0&0\mat = Q^{-1}~J~Q^{\top -1}, $$
with $$ Q=D\MAT{4} 1&0&0&0\\ 0&1&0&0\\ 1&-2\tilde{\bf
s}_1&\tilde{\bf s}_2&0\\ 0&-3&0&\tilde{\bf s}_2 \mat
$$ where $$ D=\diag\left(\frac{1}{\sqrt{\tilde{\bf
s}_2}},\frac{1}{\sqrt{\tilde{\bf
s}_2}},\frac{1}{\sqrt{3\tilde{\bf s}_2}},
\frac{1}{\sqrt{3\tilde{\bf s}_2}}\right). $$ Therefore
\bean L&=&Q~\Lambda~Q^{-1}\\ &=&\frac{1}{\tilde{\bf
s}_2^2}\MAT{4} 0&1&0&0\\
\\
2\tilde{\bf s}_1&4(\tilde{\bf s}_2-\tilde{\bf
s}_1^2)&-2\sqrt{3}\tilde{\bf s}_1&0\\
\\
\frac{4}{\sqrt{3}}(\tilde{\bf s}_2-\tilde{\bf
s}_1^2)&- \frac{8\tilde{\bf
s}_1}{\sqrt{3}}(2\tilde{\bf s}_2-\tilde{\bf
s}_1^2)&-4(\tilde{\bf s}_2-\tilde{\bf s}_1^2)&1\\
\\
-\frac{6}{\sqrt{3}}\tilde{\bf s}_1&-\frac{12}{\sqrt{3}}(\tilde{\bf
s}_2-\tilde{\bf s}^2_1)&6\tilde{\bf s}_1&0 \mat , \eean leads to
formula (7.0.5).
\end{example}
\newpage

\subsubsection*{Example 2: Two-column Jack
polynomials}

\begin{proposition} For even $N$,
choosing\footnote{$(a)_k=\displaystyle{\frac{\Gamma(a+k)}{\Gamma(a)}}=a(a+1)...(a+k-
1)$}
\be
 \left\{\begin{array}{ll}
b_0=...=b_{\frac{p}{2}-1}=0\\
\\
b_{\frac{p}{2}+k}=\frac{(1-\al)_k(p+1)_k}{k!(\al+p+1)_k},\mbox{~~for~~}
k=0,...,\frac{N-2-p}{2},
  \end{array}\right.
\label{a}\ee
 one finds the most general two-row Jack polynomial for $\tau_2$,
 for arbitrary $\al$,
 \bea
 \tau_2(t)
  &=&pf~m_2(t)
   \nonumber\\
&&\nonumber\\&=&
 J^{(1/\al)}_{\left(\frac{N+p-2}{2},\frac{N-p-2}{2}\right)}
 \left({t}/{\al} \right)\nonumber\\
 & &
 \nonumber\\
&=&
 c\oint\frac{dx}{2\pi i}\frac{dy}{2\pi
i}\frac{(y-x)^{2\al}}{(xy)^{\al+\frac{N}{2}}}
 e^{\sum_1^{\iy}t_i(x^i+y^i)}\left( \frac{x}{y}\right)^{p/2}
\!\!\!
\,_2F_1(\al,-p;1-\al-p;\frac{y}{x})\nonumber\\
 & & \nonumber\\\label{b}
 \eea
%
%
and for general $\ell\geq 2$,
  \bea \tau_{\ell}(t)&=& \frac{2c}{\ell
!!}\oint\frac{(z_2-z_1)^{2\al-1}}
 {z_2(z_1z_2)^{\al-1}}\left(\frac{z_1}{z_2}\right)^{p/2}
\,_2F_1\left(\al,-p;1-\al-p;\frac{z_2}{z_1}\right)
\nonumber\\
\nonumber\\
&&\frac{\displaystyle{\prod_{i=2}^{\ell/2}}\rho(z_{2i}/z_{2i-1})}{
\displaystyle{\prod_{i=1}^{\ell/2}}z_{2i-1}^{\frac{N}{2}-2i+3}z_{2i}^{\frac{N}{2
}-2i+1}} \prod_{1\leq
i<j\leq\ell}\left(1-\frac{z_i}{z_j}\right)\prod^{\ell}_{j=1}
 e^{\sum_{k=1}^{\iy} t_kz_j^k}\frac{dz_j}{2\pi i}~,
 \nonumber\\\label{c}\eea
  where
 \be
 \rho(x)=\sum_{i=0}^{\frac{N-2}{2}}b_{i}(x^{-i-1}-x^{i}).
\label{d}
 \ee
\end{proposition}

\newpage

\proof
According to a formula by Stanley \cite{Stanley}, two-column Jack
polynomials can be expressed as a linear combination of two-column
Schur polynomials. So, setting in the end $2s=N-2-p$, we have
%
%
%
%
%
 \bean
 \tau_2(t)
 &=&
  \sum_{k=0}^{\frac{N-2}{2}}b_k\gs_{\frac{N-2}{2}+k,\frac{N-2}{2}-k}(t),~\mbox{
with $b_k$ as in (\ref{a}),}\\
 &=&
 \sum_{k=\frac{p}{2}}^{\frac{N-2}{2}}\frac{(1-\al)_{k-p/2}(p+1)_{k-p/2}
(-1)^{N-2}}{(k-p/2)!(\al+p+1)_{k-p/2}} {\bf
s}_{\frac{N-2}{2}+k,\frac{N-2}{2}-k}(t)\\
 &=&
 \sum_{k=\frac{p}{2}}^{\frac{N-2}{2}}\frac{(1-\al)_{k-p/2}(p+1)_{k-p/2}}{(k-p/
2)!(\al+p+1)_{k-p/2}} {\bf
s}_{2^{\frac{N-2}{2}-k}1^{2k}}(-t)\\
&=&
 \sum_{k=0}^{\frac{N-p-2}{2}}\frac{(1-\al)_k(p+1)_k}{k!(\al+p+1)_k}{\bf
s}_{2^{\frac{N-2}{2}-k-p/2}1^{2k+p}}(-t)\\
 &=&
 \sum^s_{k=0}\frac{(1-\al)_k(p+1)_k}{k!(\al+p+1)_k}{\bf
s}_{2^{s-k}1^{2k+p}}(-t)
 \\
&=&
 J^{(\al)}_{2^s1^p}(-t)
  ~~~\mbox{(Stanley's formula)}\\
&=&
 J_{(p+s,s)}^{(1/\al)}(t/\al)
 ~~\mbox{using duality,}
 \eean
 showing that any two-row Jack polynomial can serve
 as Pfaff $\tau$-function $\tau_2$.

According to \cite{awata}, Jack polynomials also have an integral
representation, and so $\tau_2(t)$ can also be expressed as
\bean \tau_2(t)&=&J_{(p+s,s)}^{(1/\al)}(t/\al)\\
\\
&=&c'\oint\frac{dx}{2\pi ix}\frac{dy}{2\pi iy}\frac{dz}{2\pi
iz}\frac{(x-y)^{2\al}(xy)
^{-s}z^{-p}}{((x-z)(y-z))^{\al}}e^{\sum_1^{\iy}t_i(x^i+y^i)}\\
 \\
&=&c'\oint\frac{dx}{2\pi ix}\frac{dy}{2\pi
iy}(x-y)^{2\al}(xy)^{-s}e^{\sum_1^{\iy}t_i(x^i+y^i)}D_z^p((x-z)(y-z))^{-\al}\Big
|_{z=0}\\
 \\
&=&c'(\al)_p\oint\frac{dx}{2\pi ix}\frac{dy}{2\pi
iy}\frac{(x-y)^{2\al}}{(xy)^{\al+s}y^p}
e^{\sum_1^{\iy}t_i(x^i+y^i)}
 \,_2F_1\left(\al,-p;1-\al-p;\frac{y}{x}\right),
\eean
 where
 we used the identity:
\bean
 \lefteqn{D^p_z((x-z)(y-z))^{-\al}\Big|_{z=0}}\\
&=&(xy)^{-\al}D^p_z\left(\left(1-\frac{z}{x}\right)^{-\al}\left(
1-\frac{z}{y}\right)^{-\al}\right)\Bigg|_{z=0}\\
 \\
&=&(xy)^{-\al}D^p_z\left(\sum^{\iy}_{k,\ell=0}\frac{(\al)_k(\al)_{\ell}}{k!~\ell
!}\frac{z^{k+\ell}}{x^ky^{\ell}}\right)\Bigg|_{z=0}\\
 \\
&=&p!(xy)^{-\al}\sum_{k+\ell=p}\frac{(\al)_k(\al)_{\ell}}{k!~\ell!}x^{-k}y^{-\ell
}\\
 \\
&=&p!(xy)^{-\al}y^{-p}\sum^p_{k=0}\frac{(\al)_k(\al)_{p-k}}{k!(p-k)!}\left(\frac{y}{x}
\right)^k\\
\\
&=&(\al)_p(xy)^{-\al}y^{-p}\sum^p_{k=0}\frac{(\al)_k(-p)_k}{k!(1-\al-p)_k}
\left(\frac{y}{x}\right)^k  \\ &
&\hspace*{3.5cm}~\mbox{using~~}\frac{p!(\al)_{p-k}}{(p-k)!(\al)_p}=
\frac{(-p)_k}{(1-\al-p)_k}\\
\\
&=&(\al)_p(xy)^{-\al}y^{-p}\,_2F_1\left(\al,-p;1-\al-p;\frac{y}{x}\right)
.\eean
 This proves identity (\ref{b}).


Applying Theorem 1.3, we find the higher $\tau_{\ell}$'s, by
applying the integrated vertex operator
 \be
Y_{\frac{N-2}{2}-2j}(t)=
 \oiint X(t;z_{2j+2})X(t;z_{2j+1})\frac{\rho_b(z_{2j+2}/
z_{2j+1})dz_{2j+2}~dz_{2j+1}}{z_{2j+1}^2
 (z_{2j+1}z_{2j+2})^{\frac{N-2}{2}-2j}},
  \ee
for $j=1,2,\ldots,(\ell-2)/2$ to $\tau_2$ (see formula (\ref{b}));
so, one finds\footnote{replacing $x,y$ in $\tau_2$ with
$z_1,z_2$.}
\bean
 \tau_{\ell}&=&\frac{2}{\ell
!!}Y_{\frac{N}{2}-\ell+1}...Y_{\frac{N}{2}-5}
Y_{\frac{N}{2}-3}\tau_2\\
\\
&=&    \frac{2c'(\al)_p}{\ell
!!}\oint\frac{(z_2-z_1)^{2\al}}{(z_1z_2)^{\al+\frac{N}{2}}}\left(\frac{z_1}{z_
2}\right)^{p/2}
\,_2F_1\left(\al,-p;1-\al-p;\frac{z_2}{z_1}\right)\\
\\
&
&\frac{\rho(z_{\ell}/z_{\ell-1})...\rho(z_4/z_3)}{(z_3z_5...z_{\ell-1})^2
(z_3z_4)^{\frac{N}{2}-3}(z_5z_6)^{\frac{N}{2}-5}...(z_{\ell-1}z_{\ell})^{\frac{N
}{2}-\ell +1}}\\
 \\
 & &
 X(t;z_{\ell})X(t;z_{\ell-1})...X(t;z_4)X(t;z_3)e^{\sum_{1}^{\iy}
t_k(z_1^k+z_2^k)} \prod_{j=1}^{\ell}\frac{dz_j}{2\pi
i}\\
\\
&=&
 \frac{2c'(\al)_p}{\ell !!}\oint\frac{(z_2-z_1)^{2\al}z^2_1(z_1z_2)^
{N/2-1}}{(z_1z_2)^{\al+\frac{N}{2}}}\left(\frac{z_1}{z_2}\right)^{p/2}
\,_2F_1\left(\al,-p;1-\al-p;\frac{z_2}{z_1}\right)
\\
\\
&&
 \left(1-\frac{z_1}{z_2}\right)^{-1}\frac{\rho(z_{\ell}/z_{\ell-1})...\rho(z_4/z_3)}{\displaystyle{\prod_1^{\ell/2}
}z_{2i-1}^2 (z_{2i-1}z_{2i})^{\frac{N}{2}-2i+1}}\prod_{1\leq
i<j\leq\ell}\left(1-\frac{z_i}{z_j}\right)\prod^{\ell}_{j=1}e^{\sum_1
^{\iy}t_kz_j^k}\frac{dz_j}{2\pi i}\\
 \\
&=&  \frac{2c'(\al)_p}{\ell !!}\oint\frac{(z_2-z_1)^{2\al-1}}
 {z_2(z_1z_2)^{\al-1}}\left(\frac{z_1}{z_2}\right)^{p/2}
\,_2F_1\left(\al,-p;1-\al-p;\frac{z_2}{z_1}\right)\\
\\
&
&\frac{\displaystyle{\prod_{i=2}^{\ell/2}}\rho(z_{2i}/z_{2i-1})}{
\displaystyle{\prod_{i=1}^{\ell/2}}z_{2i-1}^{\frac{N}{2}-2i+3}z_{2i}^{\frac{N}{2
}-2i+1}} \prod_{1\leq
i<j\leq\ell}\left(1-\frac{z_i}{z_j}\right)\prod^{\ell}_{j=1}
 e^{\sum_{k=1}^{\iy} t_kz_j^k}\frac{dz_j}{2\pi i}~,
   \eean
establishing formula (\ref{c}). \qed

\noindent{\bf Alternative formula:}  The following formula has the
advantage to be more symmetric, but the disadvantage to have many
more integrations:
$$
\tau_{\ell}(t)=\oint\prod^{\ell}_{i=1}
 \prod^i_{j=1}\frac{dz_j^{(i)}} {z_j^{(i)}}
 \prod^{\ell}_{i=1}e^{\sum_1^{\iy}t_k (z^{(\ell)}_i)^{-k}}
 \frac{\displaystyle{\prod^{\ell}_{k=1}
  \prod_{{1\leq i,j\leq k}\atop{i\neq j}}}
  \left(1-\frac{z_i^{(k)}}{z_j^{(k)}}\right)}
 {\displaystyle{\prod_{k=1}^{\ell-1}
  \prod_{{1\leq i\leq k+1}\atop{1\leq j\leq k}}}
 \left(1-\frac{z_i^{(k+1)}}{z_j^{(k)}}\right)}K_{N,p,\ell}(Z)
$$
with
\bean
K_{N,p,\ell}&=&
 \frac{\displaystyle{\left(\prod^{\ell}_{j=1}
  z_j^{(\ell)}\right)^{\frac{N-p}{2}-1}
  \left(\prod^{\ell /2}_{j=1}z_j^{(\ell/2)}\right)^{p+1}}}
  {\displaystyle{\prod^{\ell-1}_{i=1}\prod^i_1z_j^{(i)}}}
   \\ &&
   \prod_{i=1}^{\ell/2}
   \,_2F_1
    \left(1-\al,p+1;1+\al+p;\frac{\displaystyle{\prod^i_{j=1}
z_j^{(i)}\prod_{j=1}^{\ell-i}z_j^{(\ell-i)}}}
 {\displaystyle{\prod_{j=1}^{i-1}z_j^{(i-1)}\prod_{j=1}^{\ell+i-i}
 z_j ^{(\ell+1-i)}}}\right). \eean

\end{document}